\documentclass[aps,prl,twocolumn,showpacs,superscriptaddress,nofootinbib]{revtex4}

\usepackage{amsmath}   
\usepackage{graphicx}  
\usepackage{xspace}
\usepackage{units}

\usepackage{hyperref}

\newcommand{\minerva}{MINERvA\xspace}
\newcommand{\minos}{MINOS\xspace}

\newcommand{\numu}{\ensuremath{\nu_{\mu}}\xspace}
\newcommand{\numubar}{\ensuremath{\bar{\nu}_{\mu}}\xspace}
\newcommand{\nubar}{\ensuremath{\bar{\nu}}}

\newcommand{\tcoh}{\ensuremath{\left| t\right| \xspace}}

\newcommand{\rmt}{\rm\textstyle}

\newcommand{\sizecheck}{0} 
\newcommand{\PRLsupp}{0}   
\ifnum\PRLsupp=0
  
  \newcommand{\SuppLocationEndSentence}{in the Appendix.}
\else
  
  \newcommand{\SuppLocationEndSentence}{at URL}
\fi

\newif\ifpdf
\ifx\pdfoutput\undefined
   \pdffalse
\else
   \pdfoutput=1
   \pdftrue
\fi
\ifpdf
   \usepackage{graphicx}
   \usepackage{epstopdf}
\else
   \usepackage{graphicx}
\fi

\begin{document}

\title{Measurement of Coherent Production of $\pi^\pm$ in Neutrino and Anti-Neutrino Beams on Carbon from $E_\nu$ of $1.5$ to $20$~GeV}



\newcommand{\deceased}{Deceased}


\newcommand{\Rutgers}{Rutgers, The State University of New Jersey, Piscataway, New Jersey 08854, USA}
\newcommand{\Hampton}{Hampton University, Dept. of Physics, Hampton, VA 23668, USA}
\newcommand{\Dortmund}{Institute of Physics, Dortmund University, 44221, Germany }
\newcommand{\Otterbein}{Department of Physics, Otterbein University, 1 South Grove Street, Westerville, OH, 43081 USA}
\newcommand{\JMU}{James Madison University, Harrisonburg, Virginia 22807, USA}
\newcommand{\Florida}{University of Florida, Department of Physics, Gainesville, FL 32611}
\newcommand{\UCIrvine}{Department of Physics and Astronomy, University of California, Irvine, Irvine, California 92697-4575, USA}
\newcommand{\CBPF}{Centro Brasileiro de Pesquisas F\'{i}sicas, Rua Dr. Xavier Sigaud 150, Urca, Rio de Janeiro, RJ, 22290-180, Brazil}
\newcommand{\PUCP}{Secci\'{o}n F\'{i}sica, Departamento de Ciencias, Pontificia Universidad Cat\'{o}lica del Per\'{u}, Apartado 1761, Lima, Per\'{u}}
\newcommand{\INRM}{Institute for Nuclear Research of the Russian Academy of Sciences, 117312 Moscow, Russia}
\newcommand{\Jlab}{Jefferson Lab, 12000 Jefferson Avenue, Newport News, VA 23606, USA}
\newcommand{\Pittsburgh}{Department of Physics and Astronomy, University of Pittsburgh, Pittsburgh, Pennsylvania 15260, USA}
\newcommand{\Guanajuato}{Campus Le\'{o}n y Campus Guanajuato, Universidad de Guanajuato, Lascurain de Retana No. 5, Colonia Centro. Guanajuato 36000, Guanajuato M\'{e}xico.}
\newcommand{\Athens}{Department of Physics, University of Athens, GR-15771 Athens, Greece}
\newcommand{\Tufts}{Physics Department, Tufts University, Medford, Massachusetts 02155, USA}
\newcommand{\WM}{Department of Physics, College of William \& Mary, Williamsburg, Virginia 23187, USA}
\newcommand{\FNAL}{Fermi National Accelerator Laboratory, Batavia, Illinois 60510, USA}
\newcommand{\Purdue}{Department of Chemistry and Physics, Purdue University Calumet, Hammond, Indiana 46323, USA}
\newcommand{\MCLA}{Massachusetts College of Liberal Arts, 375 Church Street, North Adams, MA 01247}
\newcommand{\UMD}{Department of Physics, University of Minnesota -- Duluth, Duluth, Minnesota 55812, USA}
\newcommand{\Northwestern}{Northwestern University, Evanston, Illinois 60208}
\newcommand{\UNI}{Universidad Nacional de Ingenier\'{i}a, Apartado 31139, Lima, Per\'{u}}
\newcommand{\Rochester}{University of Rochester, Rochester, New York 14627 USA}
\newcommand{\Austin}{Department of Physics, University of Texas, 1 University Station, Austin, Texas 78712, USA}
\newcommand{\USM}{Departamento de F\'{i}sica, Universidad T\'{e}cnica Federico Santa Mar\'{i}a, Avenida Espa\~{n}a 1680 Casilla 110-V, Valpara\'{i}so, Chile}
\newcommand{\Geneva}{University of Geneva, 1211 Geneva 4, Switzerland}
\newcommand{\Chicago}{Enrico Fermi Institute, University of Chicago, Chicago, IL 60637 USA}
\newcommand{\hired}{}
\newcommand{\bmeThanks}{now at SLAC National Accelerator Laboratory, Stanford, California 94309 USA}
\newcommand{\ticeThanks}{now at Argonne National Laboratory, Argonne, IL 60439, USA }

\author{A.~Higuera}                       \affiliation{\Rochester}\affiliation{\Guanajuato}
\author{A.~Mislivec}                       \affiliation{\Rochester}
\author{L.~Aliaga}                        \affiliation{\WM}  \affiliation{\PUCP}
\author{O.~Altinok}                       \affiliation{\Tufts}
\author{A.~Bercellie}                     \affiliation{\Rochester}
\author{M.~Betancourt}                    \affiliation{\FNAL}
\author{A.~Bodek}                         \affiliation{\Rochester}
\author{A.~Bravar}                        \affiliation{\Geneva}
\author{W.K.~Brooks}                      \affiliation{\USM}
\author{H.~Budd}                          \affiliation{\Rochester}
\author{A.~Butkevich}                     \affiliation{\INRM}
\author{M.F.~Carneiro}                    \affiliation{\CBPF}
\author{C.M.~Castromonte}                 \affiliation{\CBPF}
\author{M.E.~Christy}                     \affiliation{\Hampton}
\author{J.~Chvojka}                       \affiliation{\Rochester}
\author{H.~da~Motta}                      \affiliation{\CBPF}
\author{J.~Devan}                         \affiliation{\WM}
\author{S.A.~Dytman}                      \affiliation{\Pittsburgh}
\author{G.A.~D\'{i}az~}                   \affiliation{\PUCP}
\author{B.~Eberly}\thanks{\bmeThanks}   \affiliation{\Pittsburgh}
\author{J.~Felix}                         \affiliation{\Guanajuato}
\author{L.~Fields}                        \affiliation{\Northwestern}
\author{R.~Fine}                          \affiliation{\Rochester}
\author{G.A.~Fiorentini}                  \affiliation{\CBPF}
\author{H.~Gallagher}                     \affiliation{\Tufts}
\author{A.~Gomez}                         \affiliation{\Rochester}
\author{R.~Gran}                          \affiliation{\UMD}
\author{D.A.~Harris}                      \affiliation{\FNAL}
\author{K.~Hurtado}                       \affiliation{\CBPF}  \affiliation{\UNI}
\author{J.~Kleykamp}                      \affiliation{\Rochester}
\author{M.~Kordosky}                      \affiliation{\WM}
\author{T.~Le}                            \affiliation{\Rutgers}
\author{E.~Maher}                         \affiliation{\MCLA}
\author{S.~Manly}                         \affiliation{\Rochester}
\author{W.A.~Mann}                        \affiliation{\Tufts}
\author{C.M.~Marshall}                    \affiliation{\Rochester}
\author{D.A.~Martinez~Caicedo}            \affiliation{\CBPF} \affiliation{\FNAL}
\author{K.S.~McFarland}                   \affiliation{\Rochester}  \affiliation{\FNAL}
\author{C.L.~McGivern}                    \affiliation{\Pittsburgh}
\author{A.M.~McGowan}                     \affiliation{\Rochester}
\author{B.~Messerly}                      \affiliation{\Pittsburgh}
\author{J.~Miller}                        \affiliation{\USM}
\author{J.G.~Morf\'{i}n}                  \affiliation{\FNAL}
\author{J.~Mousseau}                      \affiliation{\Florida}
\author{T.~Muhlbeier}                     \affiliation{\CBPF}
\author{D.~Naples}                        \affiliation{\Pittsburgh}
\author{J.K.~Nelson}                      \affiliation{\WM}
\author{A.~Norrick}                       \affiliation{\WM}
\author{J.~Osta}                          \affiliation{\FNAL}
\author{J.L.~Palomino}                    \affiliation{\CBPF}
\author{V.~Paolone}                       \affiliation{\Pittsburgh}
\author{J.~Park}                          \affiliation{\Rochester}
\author{C.E.~Patrick}                     \affiliation{\Northwestern}
\author{G.N.~Perdue}                      \affiliation{\FNAL}  \affiliation{\Rochester}
\author{R.D.~Ransome}                     \affiliation{\Rutgers}
\author{H.~Ray}                           \affiliation{\Florida}
\author{L.~Ren}                           \affiliation{\Pittsburgh}
\author{P.A.~Rodrigues}                   \affiliation{\Rochester}
\author{D.~Ruterbories}                                 \affiliation{\Rochester}
\author{H.~Schellman}                     \affiliation{\Northwestern}
\author{D.W.~Schmitz}                     \affiliation{\Chicago}  \affiliation{\FNAL}
\author{F.D.~Snider}                      \affiliation{\FNAL}
\author{C.J.~Solano~Salinas}              \affiliation{\UNI}
\author{N.~Tagg}                          \affiliation{\Otterbein}
\author{B.G.~Tice}\thanks{\ticeThanks}    \affiliation{\Rutgers}
\author{E.~Valencia}                      \affiliation{\Guanajuato}
\author{T.~Walton}                        \affiliation{\Hampton}
\author{J.~Wolcott}                       \affiliation{\Rochester}
\author{M.Wospakrik}                      \affiliation{\Florida}
\author{G.~Zavala}                        \affiliation{\Guanajuato}
\author{D.~Zhang}                         \affiliation{\WM}
\author{B.P.Ziemer}                       \affiliation{\UCIrvine}
%
\collaboration{The \minerva  Collaboration}\ \noaffiliation

\date{\today}

\begin{abstract}
Neutrino-induced coherent charged pion production on nuclei, $\stackrel{(-)}{\nu}_\mu
A\to\mu^\pm\pi^\mp A$ is a rare, inelastic interaction in which a
small squared four-momentum \tcoh \ is transferred to the recoil nucleus leaving
it intact in the reaction.  In the scintillator tracker of \minerva,
we remove events with evidence of particles from nuclear breakup and reconstruct \tcoh\   from the final state pion and muon. We select
low \tcoh\ events to isolate a sample rich in coherent candidates. By selecting low \tcoh\ events, we produce a model-independent measurement of the
differential cross section for coherent scattering of neutrinos and anti-neutrinos on carbon. We find poor agreement with the
predicted kinematics in neutrino generators used by current oscillation experiments.
\end{abstract}
\ifnum\sizecheck=0  
\maketitle
\fi

\begin{figure}[bt]
\centering
\ifnum\PRLsupp=0
  \includegraphics[width=0.4\columnwidth]{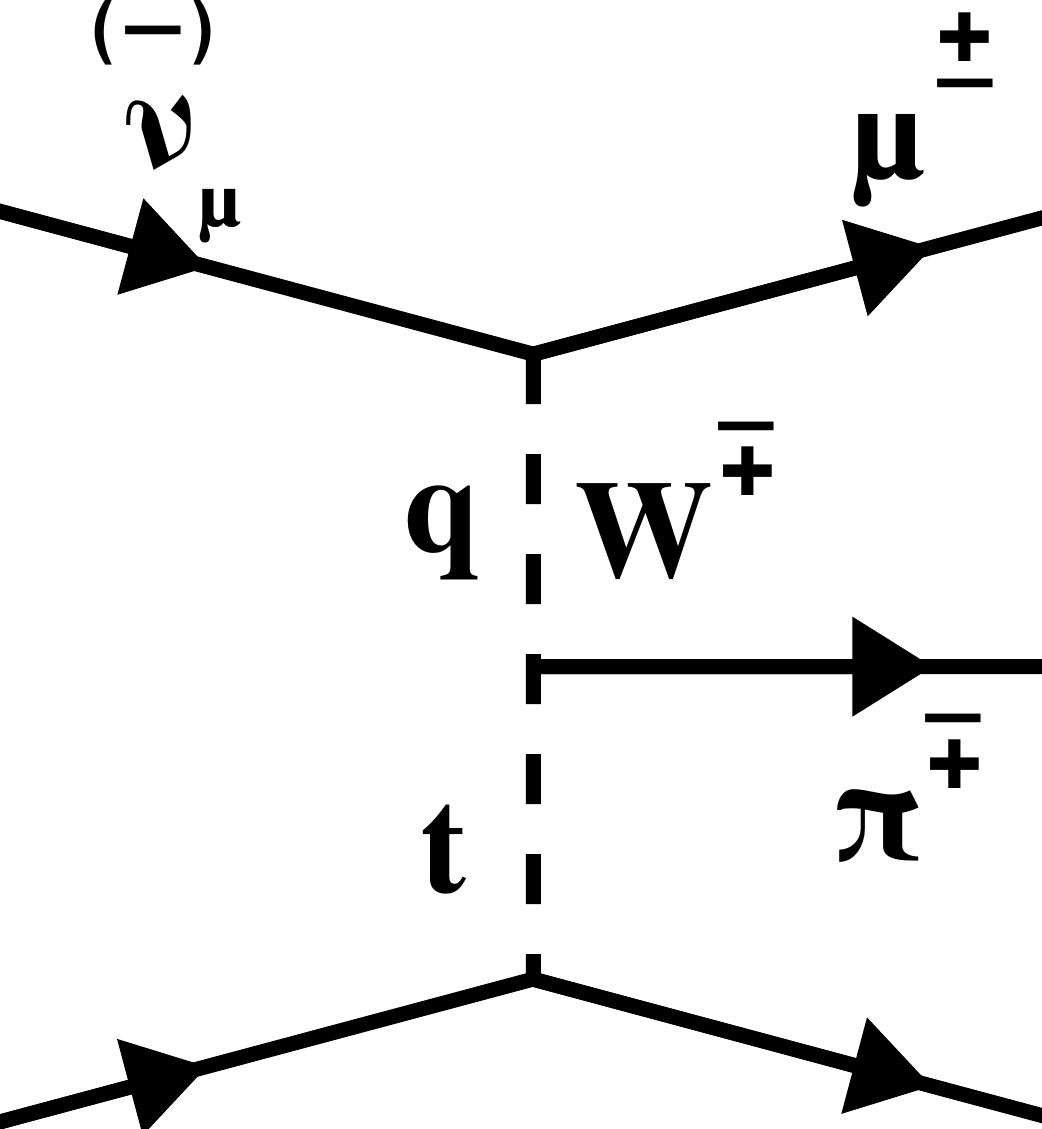}
\else
  \includegraphics[width=0.4\columnwidth]{figures/CoherentDiagram-cropped}  
\fi
\caption{Feynman diagram for coherent charged pion production}
\label{fig:feyn-coh}
\end{figure}

Coherent pion production from nuclei is an electroweak process described by the
diagram in Fig.~\ref{fig:feyn-coh} in which a virtual pion
scatters from a target nucleus that remains unchanged in its ground
state after scattering.  To achieve this coherence, the square of the
four-momentum exchanged with the nucleus must be small,
$\tcoh\lesssim \hbar^2/R^2$, where $R$ is the radius of the nucleus, and
the particle(s) exchanged can only carry vacuum quantum numbers.
Adler's theorem\cite{Adler:1964yx} provides a relationship between the
coherent scattering cross section at $Q^2\equiv-q^2=0$ and the pion-nucleus
elastic cross section\cite{Piketty:1970sq,Lackner:1979ax,Rein:1982pf}, which in the limit of
$m_\mu,m_\pi\ll E_\nu$ is
\begin{equation}\label{eqn:coh-0Q2}
\frac{d^3\sigma_{\mathrm coh}}{dQ^2dy\,d\!\tcoh}\Bigg{|}_{Q^2=0} =\frac{G_F^2}{2\pi^2}\,f_\pi^2\,\frac{1-y}{y}\,\frac{d\sigma(\pi A\to\pi A)}{d\!\tcoh},
\end{equation}
where $y$ is $E_\pi/E_\nu$ and $f_\pi$ is the pion decay constant.
The $\pi A$ elastic scattering cross section falls with increasing $\tcoh\sim e^{-\tcoh R^2/\hbar^2}$\cite{Lackner:1979ax,Rein:1982pf}.  Models must be used to extrapolate to $Q^2\neq 0$.  The model most commonly used in neutrino event
generators\cite{Andreopoulos201087, Hayato:2009zz, Casper:2002sd} is that of Rein and
Sehgal\cite{Rein:1982pf}, which assumes no vector current
and extrapolates the axial-vector current using a
multiplicative dipole form factor, $F^2_{dipole}(Q^2)=1/(1+Q^2/m_A^2)^2$, to modify Eq.~\ref{eqn:coh-0Q2}.
Other authors have
proposed alternate extrapolations to $Q^2\neq 0$\cite{Gershtein:1980vd,Belkov:1986hn,Berger:2008xs,Paschos:2009ag}.
It is also necessary to parametrize the $\pi A$ elastic scattering
cross section, and generators have varied approaches\cite{Andreopoulos201087, Hayato:2009zz, Casper:2002sd}.    
At low energies, modifications to Eq.~\ref{eqn:coh-0Q2} due to finite masses become
important, in particular $Q^2\ge m^2_\mu\frac{y}{1-y}$ and
$\tcoh\ge\left( \frac{Q^2+m_\pi^2}{2yE_\nu}\right) ^2$\cite{Rein:2006di,Higuera:2013pra}.  
An alternate approach for calculating the cross section at low
neutrino energies is to relate it to low $W$ (hadronic invariant mass) inclusive pion production\cite{Singh:2006bm,AlvarezRuso:2007tt,Amaro:2008hd,Leitner:2009ph,Hernandez:2009vm}.

Interest in coherent pion production has recently revived because
of accelerator neutrino oscillation experiments\cite{Oyama:1998bd,
Michael:2008bc, Abe:2011ks, Ayres:2004js} in which this reaction is a background to quasielastic
neutrino-nucleon interactions when a $\pi^0$ or a $\pi^\pm$ is
mistaken for an $e^\pm$ or proton, respectively.  Recently, 
low energy experiments, K2K\cite{Hasegawa:2005td} and
SciBooNE\cite{Hiraide:2008eu}, did not observe coherent $\pi^+$
production at neutrino energies $\sim 1$~GeV at the level predicted by
the Rein-Sehgal model\cite{Rein:1982pf} as then implemented in the
NEUT\cite{Hayato:2009zz} and NUANCE\cite{Casper:2002sd} event generators. There
is strong experimental evidence for coherent $\pi^0$ production at
these energies\cite{AguilarArevalo:2008xs,Kurimoto:2010rc}.

In this letter, we identify a sample of coherent
$\pi^\pm$ candidates from neutrino and anti-neutrino beams on a
scintillator (primarily CH) target by reconstructing the final state
$\mu^\mp$ and $\pi^\pm$, allowing only minimal additional energy near the
neutrino interaction vertex and requiring small \tcoh\ as a signature of the coherent reaction.  Non-coherent backgrounds are constrained with a sideband with 
high \tcoh.  In contrast to other low energy
measurements\cite{Hasegawa:2005td, Hiraide:2008eu, AguilarArevalo:2008xs,Kurimoto:2010rc, T2K-INGRID-coh,Acciarri:2014eit}
which rely on selection in the pion kinematics or in $Q^2$, this
approach uses only model-independent characteristics of coherent pion
production and therefore allows a measurement of the distribution of pion
energies and angles in coherent reactions to test the models.


The \minerva experiment studies neutrinos produced in the NuMI
beamline\cite{Anderson:1998zza}. A beam of \unit[120]{GeV} protons 
strike a graphite target, and charged mesons are focused by two magnetic horns into a \unit[675]{m} 
helium-filled decay pipe.  The horns focus positive (negative)
mesons, resulting in a $\nu_\mu$ ($\nubar_\mu$) enriched beam with a peak neutrino energy of $3.5$~GeV.
This analysis uses data taken between October 2009 and
April 2012 with $3.05\times 10^{20}$ POT (protons on target) in $\nu_\mu$ mode and $2.01\times 10^{20}$ POT in $\nubar_\mu$ mode.   

The neutrino beam is simulated in a Geant4-based
model\cite{Agostinelli2003250,1610988} constrained to
reproduce hadron production measurements on carbon by the NA49 and MIPP experiments\cite{Alt:2006fr,Lebedev:2007zz}.  
Hadronic interactions not constrained by the NA49 or MIPP data are predicted
using the FTFP hadron shower model\footnote{{F}TFP shower model in
Geant4 version 9.2 patch 03.}.  The uncertainty on the prediction of
the neutrino flux is set by the precision in these hadron production
measurements, uncertainties in the beam line focusing system and
alignment\cite{Pavlovic:2008zz}, and comparisons between different
hadron production models in regions not covered by the NA49 or MIPP
data.

The \minerva detector consists of a core of scintillator
strips surrounded by electromagnetic and hadronic calorimeters on the
sides and downstream end of the detector\footnote{The \minerva
scintillator tracking region is 95\% CH and 5\% other materials by
weight.}\cite{minerva_nim}.  The triangular $3.4\times1.7$~cm$^2$ strips are perpendicular to the
z-axis\footnote{The y-axis points along the zenith and the beam is directed downward by \unit[58]{mrad} in the y-z plane.} and are arranged in hexagonal planes.
Three plane orientations, $0^\circ$ and $\pm 60^\circ$
rotations around the z-axis, enable reconstruction of the neutrino
interaction point and the tracks of outgoing charged particles in three dimensions. The \unit[3.0]{ns}
timing resolution per plane allows separation of multiple interactions within a
single beam spill.  
\minerva is located \unit[2]{m} upstream of the MINOS near detector, 
a magnetized iron spectrometer\cite{Michael:2008bc} which is  used
in this analysis to reconstruct the momentum and charge of $\mu^\pm$.
The \minerva detector's response is simulated by a tuned 
Geant4-based\cite{Agostinelli2003250,1610988} program.  
The energy scale of the detector is set by ensuring
that both the photostatistics and the reconstructed energy deposited by
momentum-analyzed through-going muons agree in data and simulation.
The calorimetric constants used to reconstruct the energy of
$\pi^\pm$ showers and the corrections for passive material are
determined from the simulation\cite{minerva_nim}.  

To estimate backgrounds, neutrino interactions are simulated using the GENIE 2.6.2 neutrino
event generator\cite{Andreopoulos201087}.  For quasielastic
interactions, the cross section is given by the Llewellyn Smith
formalism\cite{LlewellynSmith:1971zm}.  Vector form factors come from
fits to electron scattering data\cite{Bradford:2006yz}; the axial
form factor used is a dipole with an axial mass ($M_A$) of
0.99~GeV$/$c$^2$, consistent with deuterium
measurements\cite{Bodek:2007vi,Kuzmin:2007kr}, and sub-leading form
factors are assumed from 
PCAC
or exact G-parity
symmetry\cite{Day:2012gb}.  The nuclear model is the relativistic
Fermi gas (RFG) with a Fermi momentum of
$221$~MeV$/$c and an extension to higher nucleon momenta due to short-range
correlations\cite{Bodek:1980ar,Bodek:1981wr}. Inelastic, low $W$ 
reactions are simulated with a tuned model of discrete baryon resonance
production\cite{Rein:1980wg}, and the transition to deep inelastic
scattering is simulated using the Bodek-Yang
model\cite{Bodek:2004pc}. 
Hadronization at higher energies is simulated with the AGKY model\cite{Yang:2009zx} which is based on the gradual transition
from KNO scaling to the LUND string model of PYTHIA with increasing $W$.
Final state interactions, in which hadrons
interact within the target nucleus, are modeled using the INTRANUKE
package\cite{Andreopoulos201087}.  
Uncertainties in the parameters of these models are assigned based on either measurement uncertainties
from data or to cover differences between external datasets and GENIE's model.


The \minerva\ detector\cite{minerva_nim} 
 records the energy and time of energy
depositions (hits) in each scintillator strip.  Hits are first grouped
in time and then clusters of energy are formed by spatially grouping
the hits in each scintillator plane.  Clusters with energy
$>\unit[1]{MeV}$ are then matched among the three views to create a
track.  The  $\mu^\pm$ candidate is a track that exits
the back of \minerva matching a track of the expected charge entering the
front of \minos. The most upstream cluster on the muon track is taken to be the
interaction vertex.  The resolution of each track cluster is 2.7~mm and the
angular resolution of the muon track is better than
10~mrad
in each view.  The reconstruction of the muon in the MINOS spectrometer
gives a typical muon momentum resolution of $11\%$.  Event pile-up
causes a decrease in the muon track reconstruction efficiency which was studied in both \minerva and \minos by projecting tracks found in
one detector to the other and measuring the misreconstruction
rate. This results in a $-7.8$\% ($-4.6$\%) correction to the simulated
efficiency for muons below (above) 3 GeV/c.

The interaction vertex is restricted to be
within the central 108 planes of the scintillator tracking region and
no closer than \unit[22]{cm} to any edge of the planes. These
requirements define a region with a mass of \unit[5.47]{metric tons}.
In the anti-neutrino exposure, $45\%$ of the POT were taken during the construction of the \minerva detector and therefore only used a fraction of the downstream tracker, with a fiducial volume of 56 planes and a mass of \unit[2.84]{metric tons}.

Charged $\pi^\pm$ reconstruction requires a second track originating
from the vertex.  The angular resolution on this shorter
track has a narrow central distribution with a full width at half maximum of $17$~mrad;
however the distribution has long tails due to pion scattering in the
scintillator and has an RMS resolution of $160$~mrad in each view.  
For the neutrino beam, in which CCQE events with a proton misidentified as a $\pi^+$ are a background, $dE/dx$
along the track is required to be inconsistent with that expected from
a proton ranging out in the detector.  This cut removes 64\% of
protons in the simulation while retaining 95\% of $\pi^\pm$.  The
energy of the charged pion is reconstructed calorimetrically with a fractional resolution of 
18\%+8\%/$\sqrt{E_\pi/{\rmt GeV}}$, and it is this
resolution that dominates the experimental resolution
on \tcoh.  From the measured muon and pion energies and directions,
\begin{eqnarray}
\tcoh&=&\left| \left( p_{\nu}-p_{\mu}-p_{\pi}\right) ^2\right| \nonumber\\
&\approx&\left( \sum_{i=\mu,\pi} E_i-p_{i,L}\right)
^2+\left| \sum_{i=\mu,\pi}\vec{p}_{i,T}\right| ^2, 
\end{eqnarray} 
where the approximation made is that zero energy is transferred to the
recoil nucleus and where $\vec{p}_T$ and $p_L$ are the transverse and
longitudinal momenta with respect to the known neutrino beam direction.   

By definition, a coherent reaction produces a $\mu^\mp$, a $\pi^\pm$, and
nothing else originating from the interaction vertex.  Vertex energy is defined as all energy deposited within five planes of the plane in which
the neutrino interacts in strips within $20$~cm of the interaction
vertex.  Energy deposited on the muon and pion tracks is corrected
for path length in the bars.  For coherent events, this results in a vertex energy
of $50$~MeV with an RMS spread of $10$~MeV due to fluctuations in
energy deposited by the muon and pion.  Background processes typically
 leave significantly more energy in this region, and this analysis
requires the reconstructed vertex energy to be between $30$ and
$70$~MeV.  This requirement removes 85(86)\% of the predicted
background in the $\numu$($\numubar$) measurement and keeps
87.0(86.7)\% of the coherent pion events.

\begin{figure}[tb]
\centering
  \mbox{\includegraphics[width=0.49\columnwidth]{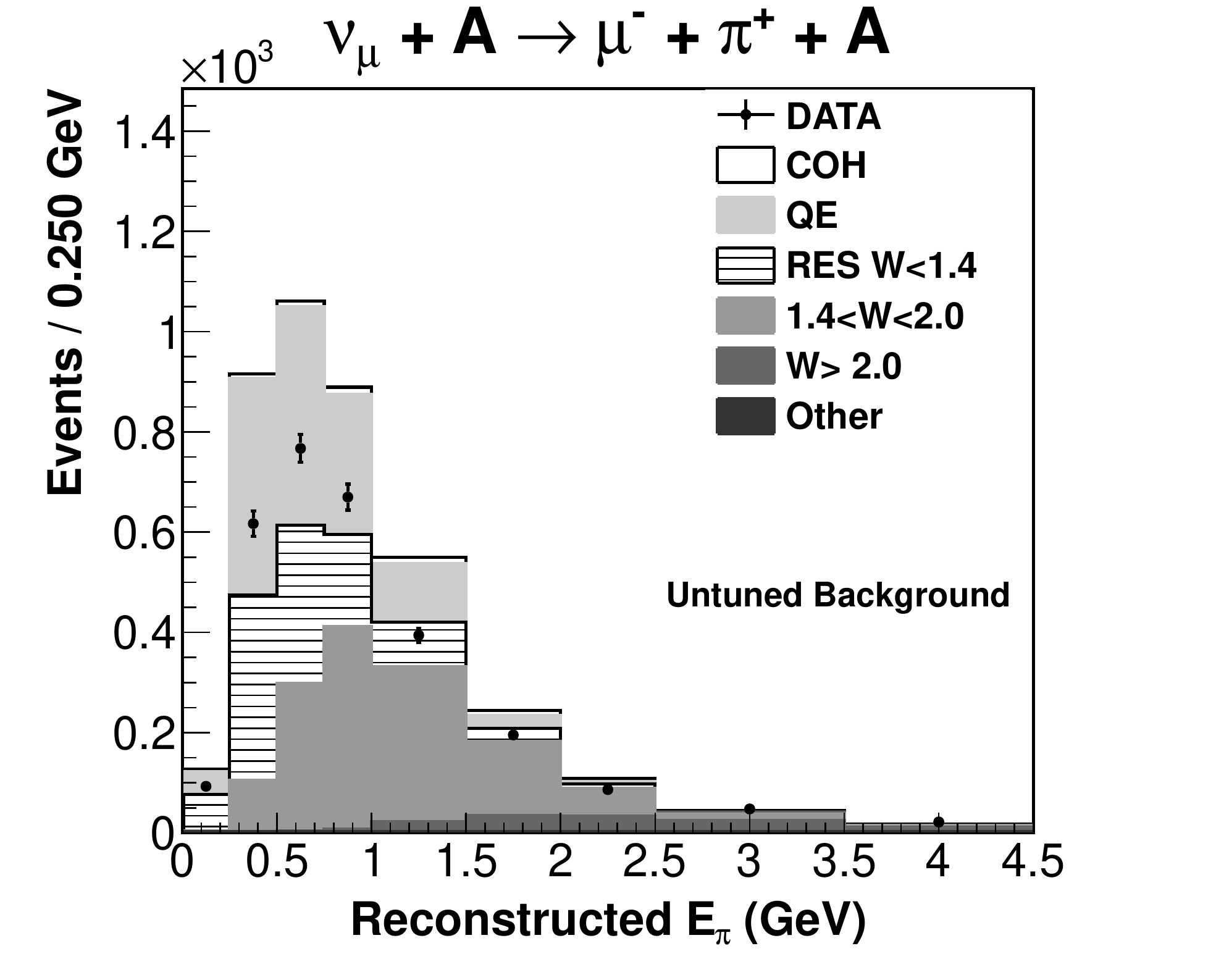}
  \includegraphics[width=0.49\columnwidth]{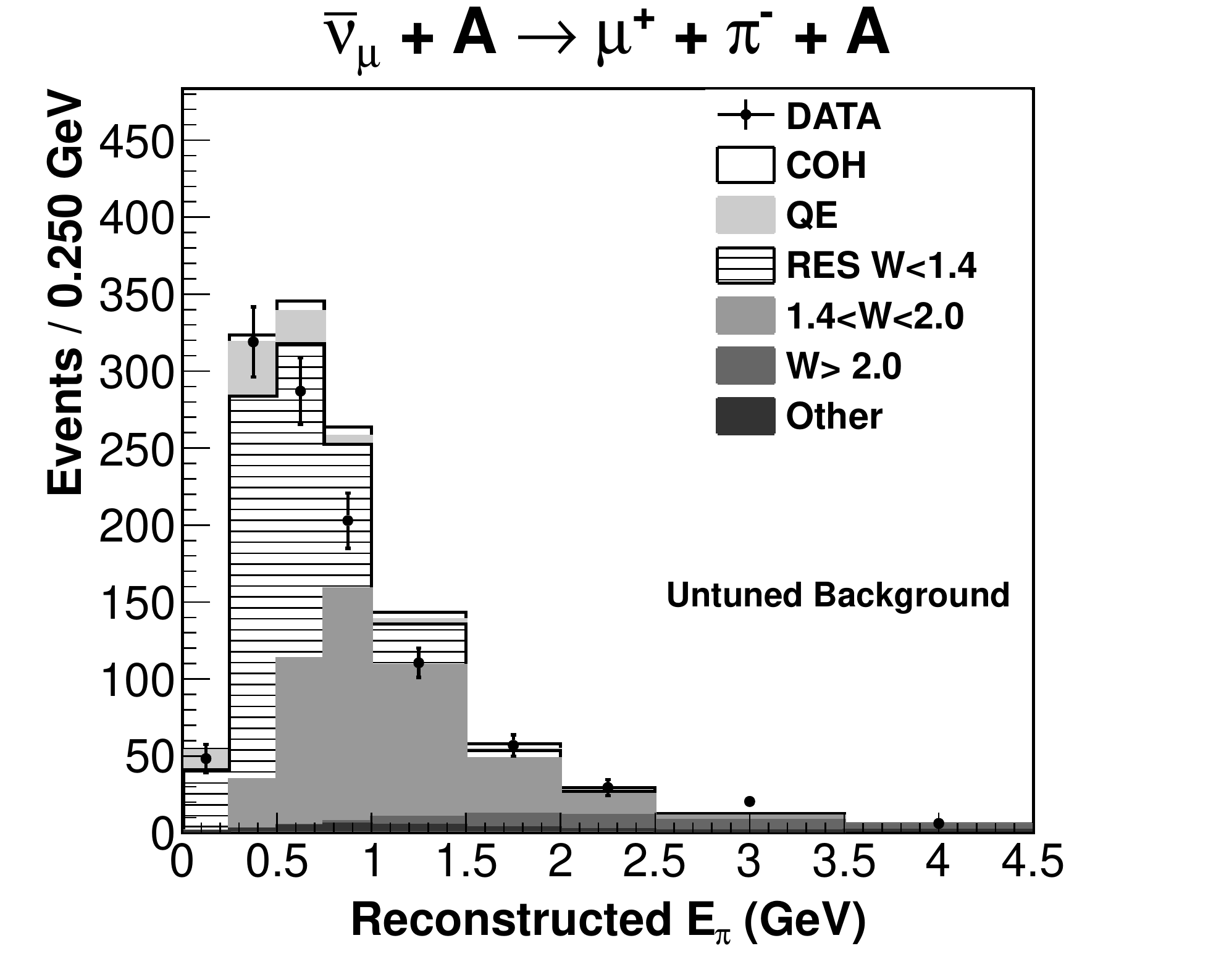}}
  \mbox{\includegraphics[width=0.49\columnwidth]{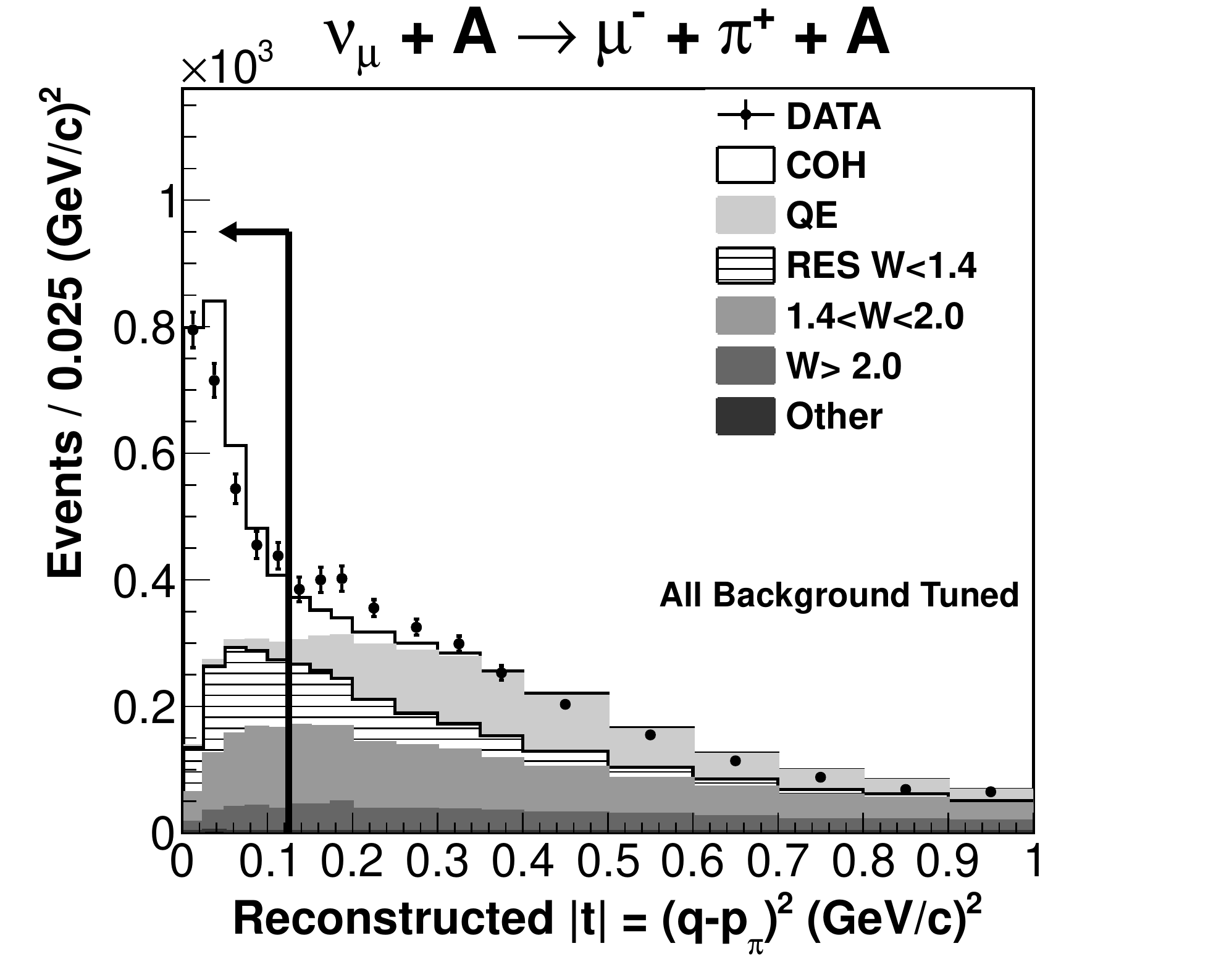}
  \includegraphics[width=0.49\columnwidth]{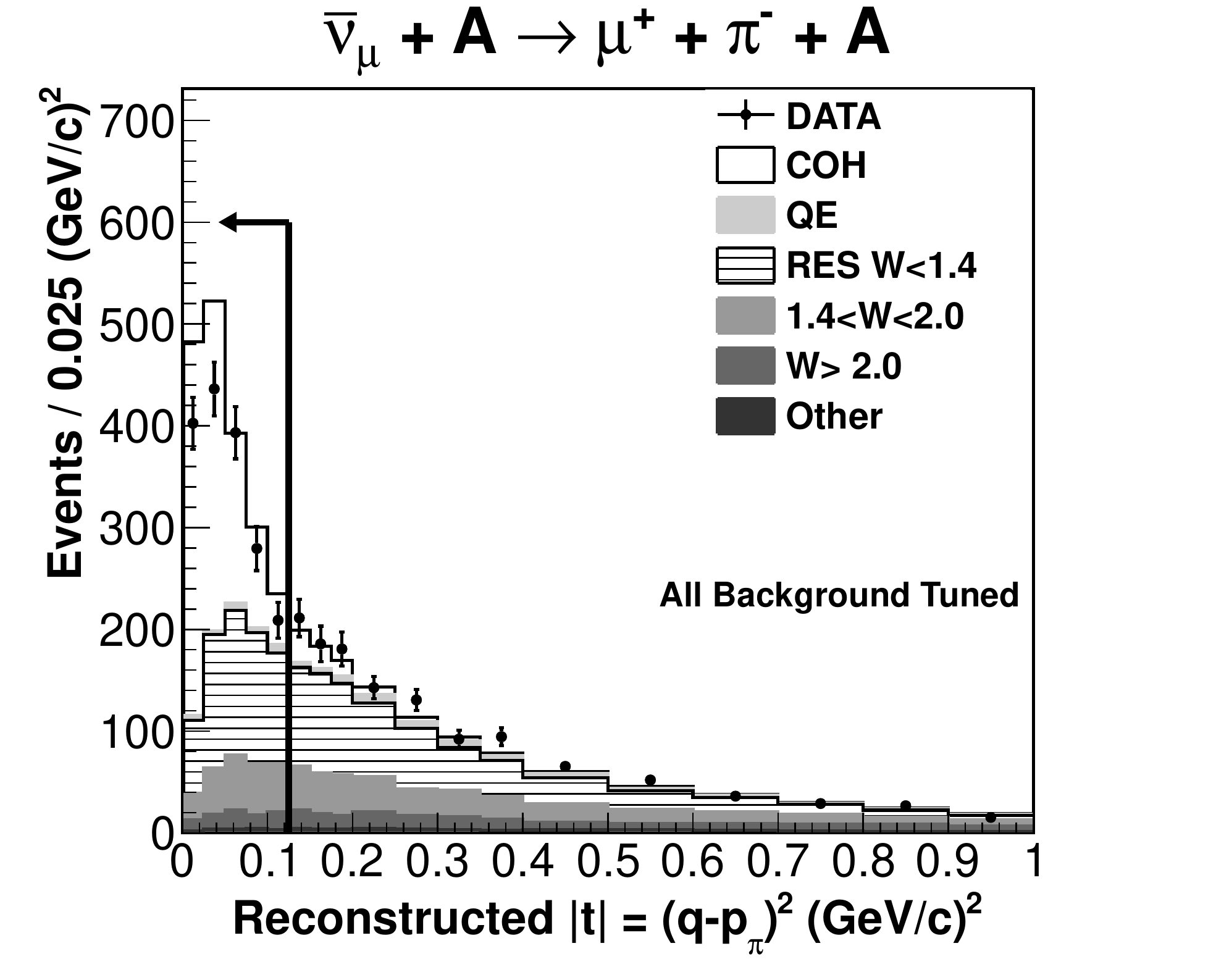}}
\caption{(Top) pion energy distribution for events in the $0.2<\tcoh<0.6$~(GeV/c)$^2$ sideband,  $\numu$ (left) and
  $\numubar$ (right) beams 
  and (bottom) reconstructed $\tcoh$ after background tuning using the sideband.  The signal distribution in $\tcoh$ peaks near zero with its shape dominated by detector resolution.}
\label{fig:tdists}
\end{figure}

As shown in Fig.~\ref{fig:tdists} (top), after the vertex energy requirement, the simulation exceeds the background at high \tcoh.  
The incoherent background components are divided into categories based on $W$, and scale factors for the background are estimated by fits to
the distributions of $\pi^\pm$ energies
for events with $0.2<\tcoh<0.6$~(GeV/c)$^2$. As shown in Table~\ref{tab:sidebandScaleFactors} the fit reduces the predicted background, particularly at low $W$. The reconstructed \tcoh\ distribution after background tuning is shown in Fig.~\ref{fig:tdists} (bottom), and a significant excess of low \tcoh\ events over the background-only prediction is observed.
The scaled background is
then subtracted from the events with $\tcoh<0.125$~(GeV/c)$^2$ to measure
the rate of coherent events in the data.  There are 1628 and 770 coherent candidates after background
subtraction in the neutrino and anti-neutrino samples respectively. 

\begingroup
\squeezetable
\begin{table}[t]
\begin{tabular}{lcc}
Source of Background & \numu & \numubar \\ \hline
Charged Current Quasielastic & $0.7\pm0.3$ & $1$ (fixed) \\
Non-quasielastic, $W<1.4$~GeV & $0.6\pm0.3$ & $0.7\pm 0.1$ \\
$1.4<W<2.0$~GeV & $0.7\pm0.1$ & $0.6\pm0.1$\\
$W>2.0$~GeV & $1.1\pm0.1$ & $1.9\pm 0.3$  
\end{tabular}
\caption{Scale factors and their statistical uncertainties determined for different background sources, grouped by hadronic invariant mass, $W$, from the high $\tcoh$ sidebands}
    \vspace{-10pt}
\label{tab:sidebandScaleFactors}
\end{table}
\endgroup 

The cross section is
determined by $\sigma = N_{\rmt coh}/\Phi N_{^{12}C}$.
where $\Phi$ is the total flux of neutrinos incident on the detector.
Our scintillator has free protons in numbers equal to the $^{12}C$
nuclei.  GENIE does not simulate diffractive production of pions from
the free protons which might also produce events at low \tcoh. There
is no microphysical calculation of this process at our energies.  An
inclusive calculation of $\stackrel{(-)}{\nu}p\to\mu^\pm\pi^\mp p$
based on $\pi A$ elastic scattering data and the Adler
relation\cite{Adler:1964yx,Kopeliovich:private}, shows a modest
low \tcoh\ enhancement not seen in GENIE which falls exponentially with \tcoh.  This difference does not identify diffractive
events, but instead includes all low \tcoh\ enhancements in scattering
from protons that might be in this calculation.  
Moreover, most events
with \tcoh$>$0.05~GeV$^2$ would not pass our vertex energy requirement because of the
recoiling proton's ionization.  We estimate the acceptance of these
low \tcoh\ events to be $\approx$20\% of the acceptance for coherent
events on carbon.  Based this low \tcoh\ enhancement and our
acceptance, the event rate in our data would be equivalent to 7\%(4\%) of the 
GENIE prediction for the coherent cross-section on
$^{12}$C for neutrinos(anti-neutrinos). We do not correct our result
for this possible enhancement.

\begin{figure}[tb]
\centering
\ifnum\PRLsupp=0
  \mbox{\includegraphics[width=0.49\columnwidth]{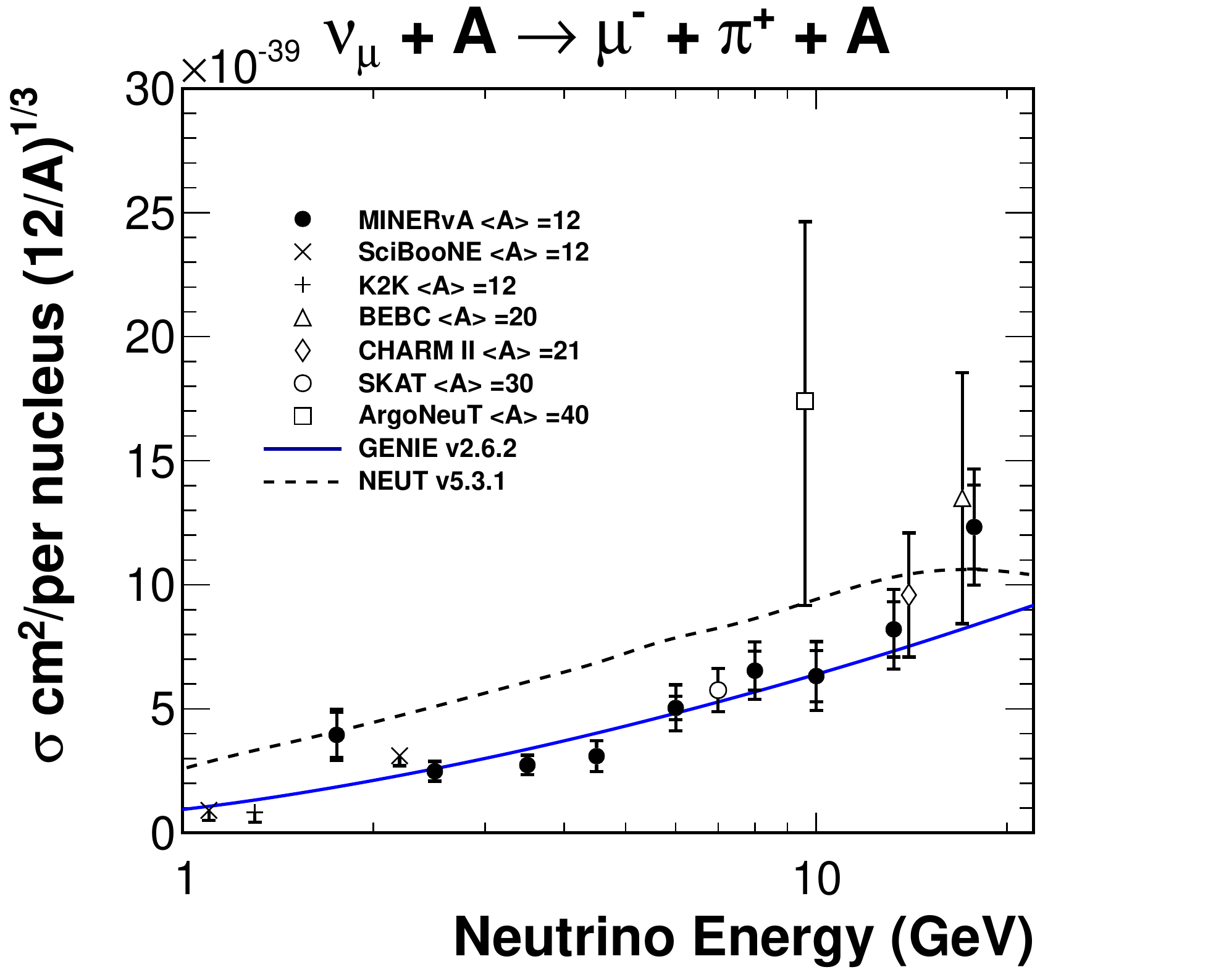}
  \includegraphics[width=0.49\columnwidth]{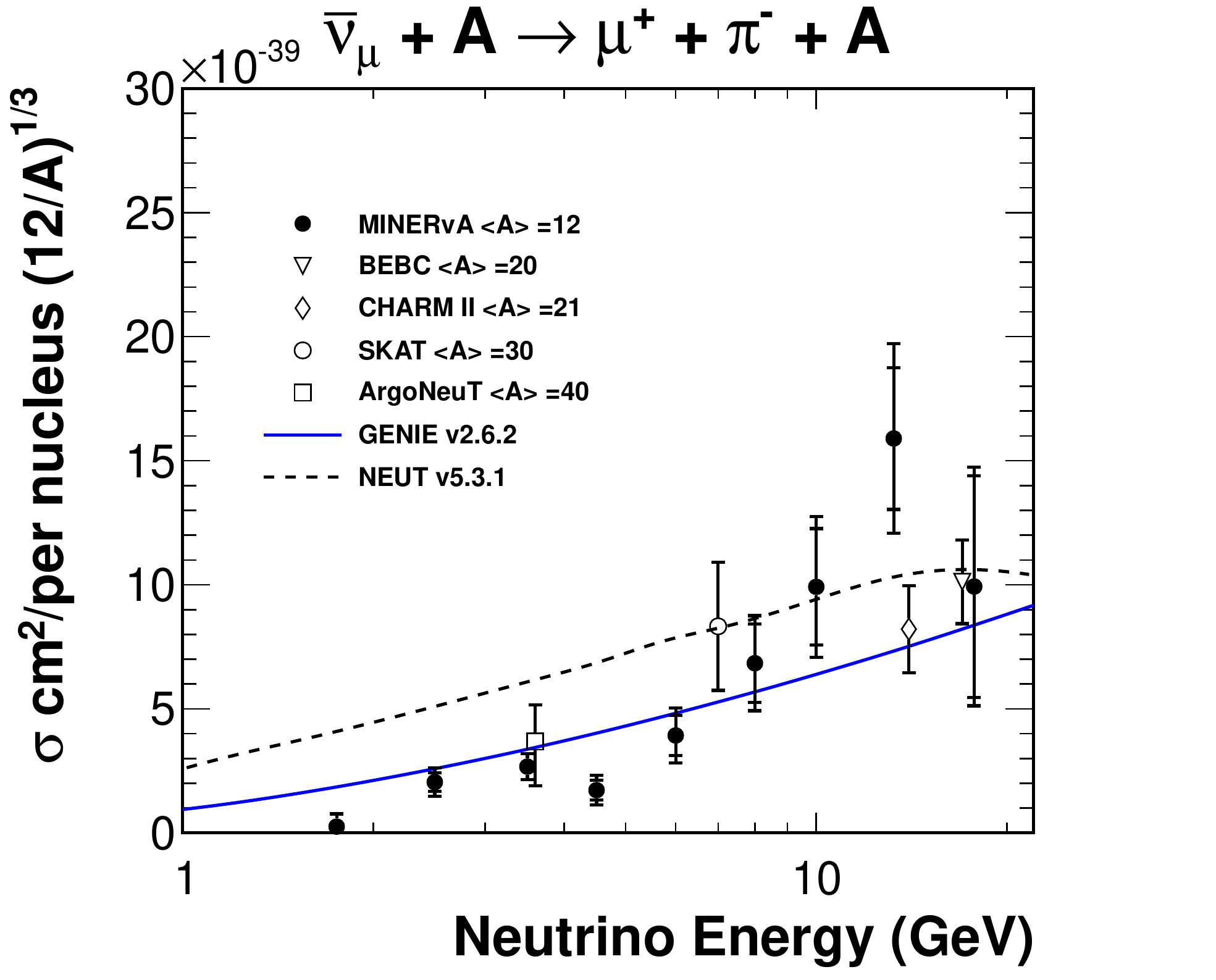}}

\else
  \mbox{\includegraphics[width=0.49\columnwidth]{figures/Neutrino_Cross_sections_witherrors.pdf}
  \includegraphics[width=0.49\columnwidth]{figures/AntiNeutrino_Cross_sections_witherrors.pdf}}

\fi
\caption{$\sigma(E_\nu)$ for neutrino (left) and anti-neutrino
  (right) coherent $\pi^\pm$ production.  The inner error bars in the cross section represent statistical uncertainties and the outer the total uncertainties; the cross section is tabulated \SuppLocationEndSentence \hfill Results from other measurements\cite{Grabosch:1985mt,Marage:1986cy,Allport:1988cq,Vilain:1993sf,Hasegawa:2005td, Hiraide:2008eu,Acciarri:2014eit} are scaled to carbon using the predicted $A^{\frac{1}{3}}$ dependence of the Rein-Sehgal model\cite{Rein:1982pf}.}
\label{fig:xsec-enu}
\end{figure}

We measure flux-averaged\footnote{The fluxes of neutrinos and anti-neutrinos in this analysis are given \SuppLocationEndSentence} cross sections of $(3.49\pm0.11{\rm\textstyle (stat)}\pm0.37{\rm\textstyle (flux)}\pm0.20{\rm\textstyle (other~sys.)})\times 10^{-39}$ and $(2.65\pm0.15{\rm\textstyle (stat)}\pm0.31{\rm\textstyle (flux)}\pm0.30{\rm\textstyle (other~sys.)})\times 10^{-39}$ cm$^2$ per $^{12}$C nucleus in the neutrino and anti-neutrino beams respectively.
 In cross-sections as a function of $E_\nu$, $E_\pi$ and $\theta_\pi$, the effect of detector resolution is accounted for
by using iterative Bayesian
unfolding\cite{D'Agostini:1994zf,Adye:2011gm}.  Figure~\ref{fig:xsec-enu} shows the measured cross sections
as a function of $E_\nu$
compared with previous measurements for $E_\nu<20$~GeV and with the
NEUT\cite{Hayato:2009zz} and GENIE\cite{Andreopoulos201087} implementations\footnote{The pion-nucleus elastic cross section correction of GENIE 2.8.0 was implemented in the simulation.}
 of Rein and
Sehgal\cite{Rein:1982pf} with lepton mass corrections\cite{Rein:2006di}.

 \begingroup
 \squeezetable
 \begin{table}
 \begin{tabular}{lccccccc}
 &$E_{\nu(\nubar)}$ (GeV) & I & II & III & IV & V  &  Total \\
 \hline
&$1.5 - 2.0$ & 0.101 & 0.041 & 0.031 & 0.017 & 0.002 & 0.115\\ 
&$2.0 - 3.0$ & 0.108 & 0.058 & 0.040 & 0.034 & 0.020 & 0.135\\ 
&$3.0 - 4.0$ & 0.099 & 0.053 & 0.041 & 0.037 & 0.027 & 0.127\\ 
&$4.0 - 5.0$ & 0.163 & 0.046 & 0.040 & 0.038 & 0.029 & 0.180\\ 
$\nu$&$5.0 - 7.0$ & 0.146 & 0.031 & 0.034 & 0.034 & 0.023 & 0.159\\ 
&$7.0 - 9.0$ & 0.118 & 0.027 & 0.034 & 0.026 & 0.016 & 0.129\\ 
&$9.0 - 11.0$ & 0.132 & 0.034 & 0.044 & 0.029 & 0.019 & 0.147\\ 
&$11.0 - 15.0$ & 0.132 & 0.022 & 0.034 & 0.021 & 0.014 & 0.140\\ 
&$15.0 - 20.0$ & 0.128 & 0.015 & 0.022 & 0.013 & 0.009 & 0.132\\ 
\hline
&$1.5 - 2.0$ & 0.216 & 0.497 & 0.309 & 0.308 & 0.111 & 0.704\\ 
&$2.0 - 3.0$ & 0.135 & 0.144 & 0.075 & 0.027 & 0.030 & 0.215\\ 
&$3.0 - 4.0$ & 0.100 & 0.095 & 0.065 & 0.021 & 0.026 & 0.156\\ 
&$4.0 - 5.0$ & 0.191 & 0.138 & 0.105 & 0.036 & 0.048 & 0.265\\ 
$\bar{\nu}$&$5.0 - 7.0$ & 0.164 & 0.065 & 0.073 & 0.025 & 0.028 & 0.194\\ 
&$7.0 - 9.0$ & 0.140 & 0.043 & 0.056 & 0.017 & 0.019 & 0.158\\ 
&$9.0 - 11.0$ & 0.148 & 0.038 & 0.048 & 0.014 & 0.015 & 0.161\\ 
&$11.0 - 15.0$ & 0.157 & 0.016 & 0.024 & 0.008 & 0.006 & 0.160\\ 
&$15.0 - 20.0$ & 0.154 & 0.052 & 0.066 & 0.017 & 0.024 & 0.178\\ 

 \hline
 \end{tabular}
 \caption{Fractional systematic uncertainties on $\sigma(E_\nu)$ and $\sigma(E_{\nubar})$  associated with Flux (I), neutrino interaction models (II), detector simulation (III), vertex energy (IV), and (V) sideband model.  The final column shows the total systematic uncertainty due to all sources.}
 \label{tab:systematics} 
 \end{table}
 \endgroup

The main sources of systematic uncertainty on the cross sections are the
flux, the background interaction model, pion interactions in the detector, muon
reconstruction, muon and hadron energy scale, vertex energy, and the
model used in the sideband constraint for the background.
These systematic uncertainties are shown in Table~\ref{tab:systematics}.
The uncertainty of hadron interactions in the detector as predicted by Geant4 on tracking and energy measurements is evaluated
by varying the pion and proton
total inelastic cross sections by $\pm$10\% and the neutron mean free path as a function of kinematic energy
by 10--25\% to span differences between Geant4 and hadron scattering data\cite{Ashery:1981tq,Allardyce:1973ce,Saunders:1996ic,Lee:2002eq,Abfalterer:2001gw,Schimmerling:1973bb,Voss:1956,Slypen:1995fm,Franz:1989cf,Tippawan:2008xk,Bevilacqua:2013rfq,Zanelli:1981zz}.

%
For muons reconstructed by range in MINOS, the muon energy scale uncertainty is
dominated by energy loss uncertainties, and
we compared range and curvature measurements to evaluate uncertainties
on reconstruction of muons by curvature in the MINOS magnetic field.  
Uncertainties in the hadron energy reconstruction result from uncertainties in the energy scale set
by muon energy deposition, material composition and dimensions, saturation of
ionization in the scintillator, and photosensor cross talk and non-linearity.
Comparisons with the test beam\cite{minerva_nim} limit the energy scale uncertainty for
pions (protons) to 5\% (3\%).  
The target mass is uncertain to $1.4$\%.

Uncertainties in predictions for the non-coherent background from the
GENIE generator are evaluated by 
varying the
underlying model tuning parameters according to their
uncertainties\cite{Andreopoulos201087}.  The most important
parameters are the normalization and axial form factor for baryon resonance production.  MINERvA's measurements of the CCQE
process\cite{nubarprl,nuprl} show that GENIE does
not model the energetic final state proton multiplicity well, which in turn means a
mismodeling of the vertex energy.  The resulting uncertainty is estimated by
turning on and off the addition of energy deposited by a 20--225 MeV final state proton to
the vertex energy of 25\% of background events with a target neutron.
Finally, after tuning the background we find remaining disagreement in the
sideband $\theta_\pi$ distribution.  This disagreement is corrected and 
the size of the correction is taken as a systematic uncertainty.
The effects of these model variations are reduced by
sideband tuning of the background.

\begin{figure}[tb]
\centering
\ifnum\PRLsupp=0
  \mbox{\includegraphics[width=0.49\columnwidth]{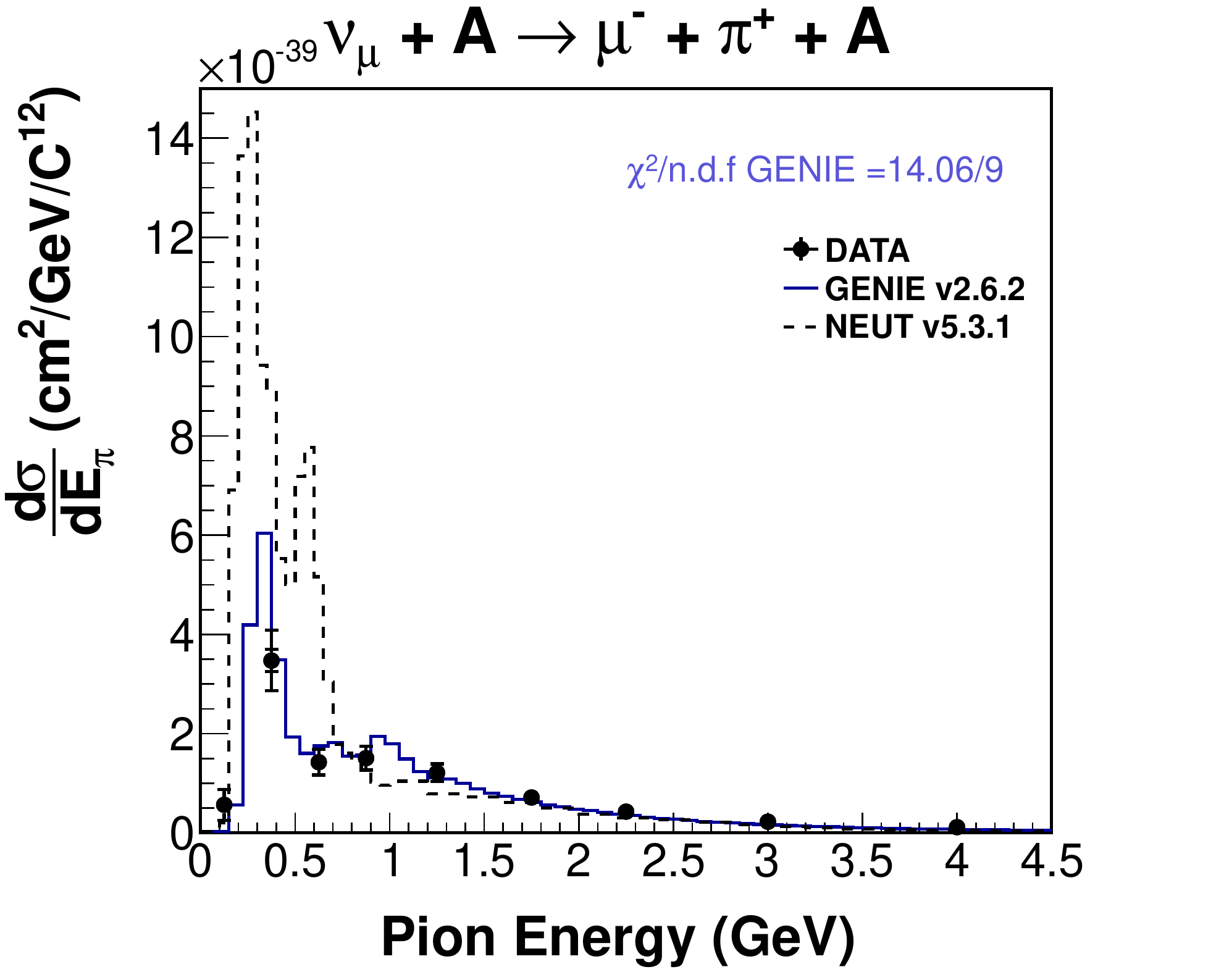}
  \includegraphics[width=0.49\columnwidth]{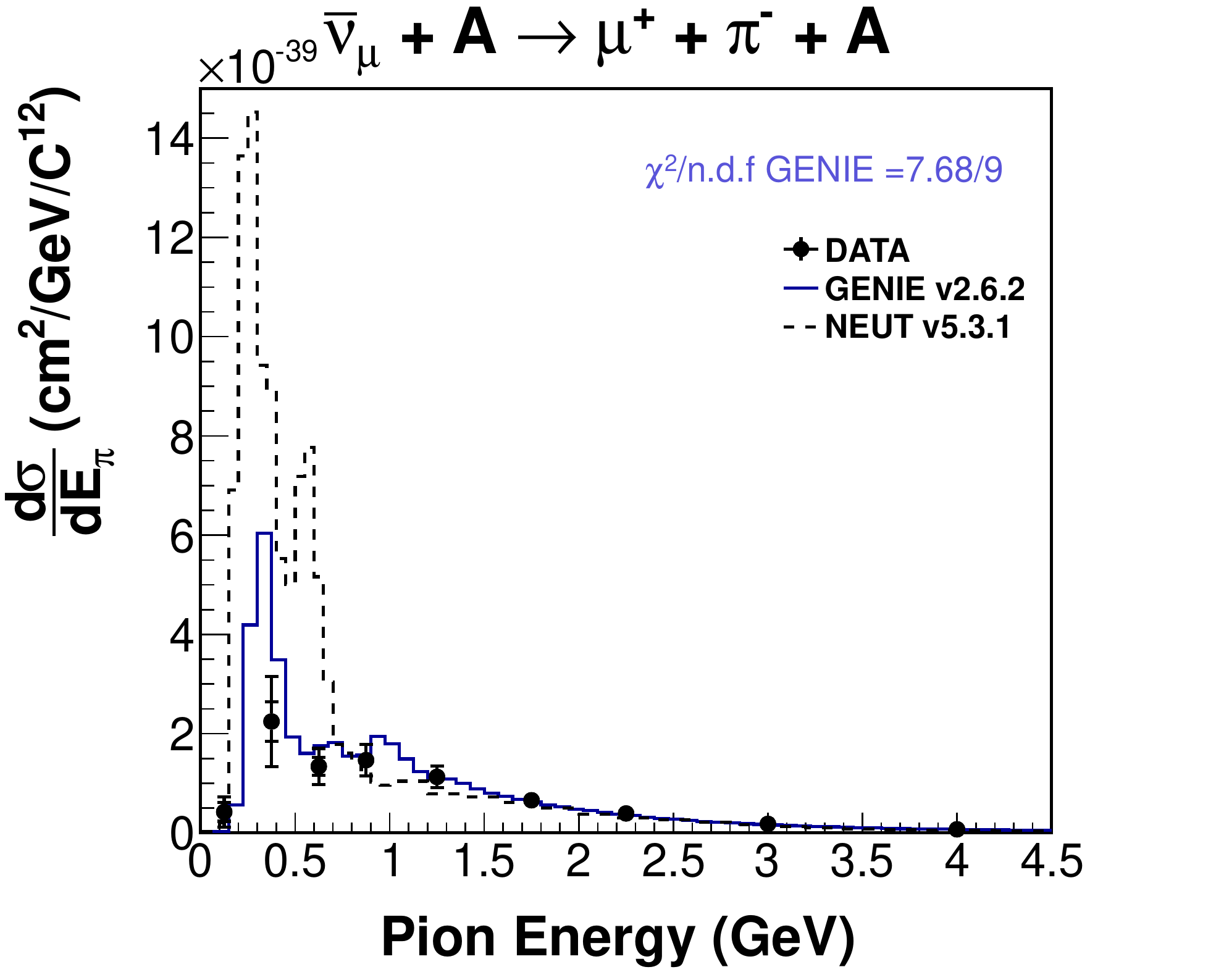}}
  \mbox{\includegraphics[width=0.49\columnwidth]{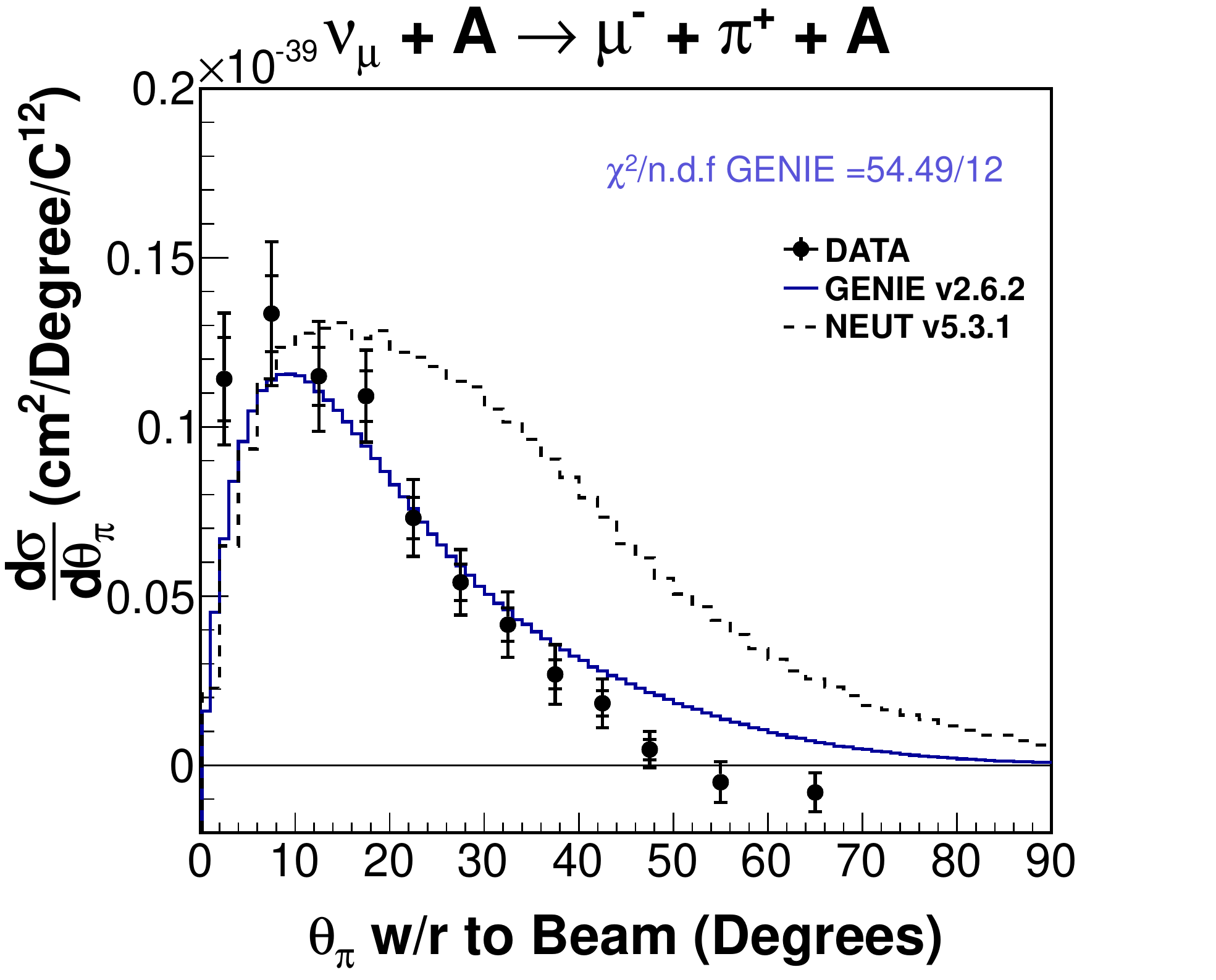}
  \includegraphics[width=0.49\columnwidth]{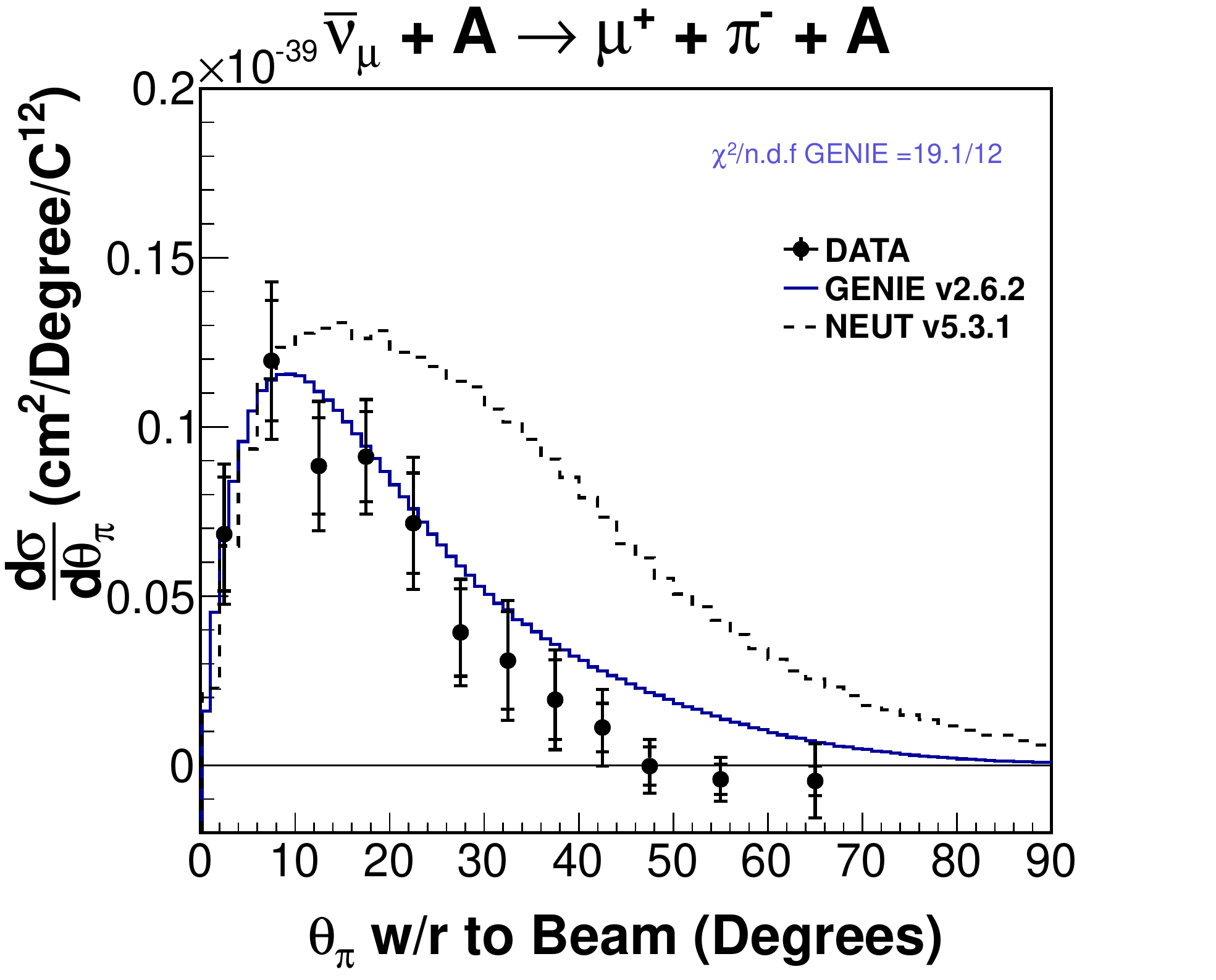}}
\else
  \mbox{\includegraphics[width=0.49\columnwidth]{figures/nu_EpiFinal_XS_pot.pdf}
  \includegraphics[width=0.49\columnwidth]{figures/nubar_EpiFinal_XS_pot.pdf}}
  \mbox{\includegraphics[width=0.49\columnwidth]{figures/nu_ThetaFinal_XS_pot.pdf}
  \includegraphics[width=0.49\columnwidth]{figures/nubar_ThetaFinal_XS_pot.pdf}}
\fi
\caption{$d\sigma/dE_\pi$ (top) and $d\sigma/d\theta_\pi$ (bottom)
  for $\numu$ (left) and
  $\numubar$ (right) with error bars as in Fig.~\ref{fig:xsec-enu} compared against predicted cross-sections from GENIE\cite{Andreopoulos201087} and NEUT\cite{Hayato:2009zz}.  These cross sections are tabulated \SuppLocationEndSentence 
}
\label{fig:xsec-pion}
\end{figure}

Figure~\ref{fig:xsec-pion} compares the flux-averaged
 differential cross sections as a function
of pion energy and angle against the Rein-Sehgal
model\cite{Rein:1982pf} as implemented in GENIE\cite{Rein:2006di,Andreopoulos201087} and NEUT\cite{Hayato:2009zz}.  Disagreement at high $\theta_\pi$ is evident in both GENIE and NEUT.  In GENIE, whose behavior is more similar to the data, the model predicts $\sim$15\% of the cross section with $\theta_\pi>45^\circ$ but there is no evidence for such events in the data.

In conclusion, the coherent production of pions on carbon nuclei for
both neutrino and anti-neutrino beams is precisely measured by
isolating a sample with no visible nuclear breakup and low \tcoh\
transferred to the nucleus.  This allows a study of produced pion
kinematics independent of the details of the signal model.  The
cross sections of the neutrino and anti-neutrino coherent pion
production are similar, indicating that the reaction is likely to be
primarily an axial vector process.  The
discrepancies observed at neutrino energies relevant for the T2K oscillation experiment\cite{Abe:2011ks} suggest
that these data should be used to revise the predictions of neutrino
interaction models used in future measurements.

\begin{acknowledgments}

This work was supported by the Fermi National Accelerator Laboratory
under US Department of Energy contract
No. DE-AC02-07CH11359 which included the \minerva construction project.
Construction support was
also granted by the United States National Science Foundation under
Grant PHY-0619727 and by the University of Rochester. Support for
participating scientists was provided by NSF and DOE (USA) by CAPES
and CNPq (Brazil), by CoNaCyT (Mexico), by CONICYT (Chile), by
CONCYTEC, DGI-PUCP and IDI/IGI-UNI (Peru), by Latin American Center for
Physics (CLAF) and by RAS and the Russian Ministry of Education and Science (Russia).  We
thank the MINOS Collaboration for use of its
near detector data. Finally, we thank the staff of
Fermilab for support of the beamline and the detector.

\end{acknowledgments}
\bibliographystyle{apsrev4-1}
\bibliography{Coherent}

\ifnum\PRLsupp=0
  \clearpage
\newcommand{\qsq}{\ensuremath{Q^2_{QE}}\xspace}
\renewcommand{\textfraction}{0.05}
\renewcommand{\topfraction}{0.95}
\renewcommand{\bottomfraction}{0.95}
\renewcommand{\floatpagefraction}{0.95}
\renewcommand{\dblfloatpagefraction}{0.95}
\renewcommand{\dbltopfraction}{0.95}
\setcounter{totalnumber}{5}
\setcounter{bottomnumber}{3}
\setcounter{topnumber}{3}
\setcounter{dbltopnumber}{3}

\onecolumngrid
\appendix{Appendix: Supplementary Material}\hfill\vspace*{4ex}\

\begingroup
\squeezetable
\begin{table}[h]
\tabcolsep=0.11cm
\begin{tabular}{c|c|c}
&  $d\sigma/dE_{\pi}\ \  (cm^{2}/GeV/C^{12})\times 10^{-39}$ & $d\sigma/dE_{\pi}\ \  (cm^{2}/GeV/C^{12})\times 10^{-39}$ \\
$E_{\pi}$ \ (GeV)& neutrinos & antineutrinos  \\
\hline

0 - 0.25 & 0.567 $\pm $0.116 $\pm$0.288 &0.423 $\pm $0.146 $\pm$0.238\\
0.25 - 0.5 & 3.473 $\pm $0.223 $\pm$0.568 &2.245 $\pm $0.313 $\pm$0.820\\
0.5 - 0.75 & 1.426 $\pm $0.105 $\pm$0.234 &1.342 $\pm $0.142 $\pm$0.315\\
0.75 - 1.0 & 1.508 $\pm $0.096 $\pm$0.222 &1.467 $\pm $0.127 $\pm$0.272\\
1.0 - 1.5 & 1.213 $\pm $0.078 $\pm$0.161 &1.129 $\pm $0.094 $\pm$0.182\\
1.5 - 2.0 & 0.714 $\pm $0.061 $\pm$0.094 &0.659 $\pm $0.073 $\pm$0.110\\
2.0 - 2.5 & 0.429 $\pm $0.047 $\pm$0.060 &0.393 $\pm $0.059 $\pm$0.074\\
2.5 - 3.5 & 0.227 $\pm $0.030 $\pm$0.034 &0.182 $\pm $0.035 $\pm$0.045\\
3.5 - 4 & 0.116 $\pm $0.019 $\pm$0.020 &0.073 $\pm $0.017 $\pm$0.020\\
 
\end{tabular}
\caption{The measurement of the differential cross section as a function of pion energy, $d\sigma/dE_\pi$, with statistical and systematic uncertainties}
\end{table}
\endgroup

\begingroup
\squeezetable
\begin{table}[h]
\tabcolsep=0.11cm
\begin{tabular}{c|c|c}
 &  $d\sigma/d\theta_{\pi}\ \  (cm^{2}/Degree/C^{12})\times 10^{-39}$ & $d\sigma/d\theta_{\pi}\ \  (cm^{2}/Degree/C^{12})\times 10^{-39}$\\ 
$\theta_{\pi}$\ (Degrees) &  neutrinos & antineutrinos\\ 
\hline
0 - 5 & 0.114 $\pm$0.012 $\pm$0.015 &0.068 $\pm$0.014 $\pm$0.012\\
5 - 10 & 0.133 $\pm$0.011 $\pm$0.018 &0.120 $\pm$0.015 $\pm$0.015\\
10 - 15 & 0.115 $\pm$0.009 $\pm$0.014 &0.088 $\pm$0.012 $\pm$0.013\\
15 - 20 & 0.109 $\pm$0.008 $\pm$0.011 &0.091 $\pm$0.011 $\pm$0.010\\
20 - 25 & 0.073 $\pm$0.006 $\pm$0.010 &0.072 $\pm$0.013 $\pm$0.013\\
25 - 30 & 0.054 $\pm$0.005 $\pm$0.008 &0.039 $\pm$0.012 $\pm$0.009\\
30 - 35 & 0.042 $\pm$0.005 $\pm$0.008 &0.031 $\pm$0.013 $\pm$0.010\\
35 - 40 & 0.027 $\pm$0.004 $\pm$0.008 &0.019 $\pm$0.011 $\pm$0.009\\
40 - 45 & 0.018 $\pm$0.004 $\pm$0.006 &0.011 $\pm$0.006 $\pm$0.009\\
45 - 50 & 0.005 $\pm$0.003 $\pm$0.004 &-0.000 $\pm$0.004 $\pm$0.006\\
50 - 60 & -0.005 $\pm$0.002 $\pm$0.006 &-0.004 $\pm$0.004 $\pm$0.005\\
60 - 70 & -0.008 $\pm$0.002 $\pm$0.005 &-0.005 $\pm$0.003 $\pm$0.010\\
 
\end{tabular}
\caption{The measurement of the differential cross section as a function of pion angle, $d\sigma/d\theta_\pi$, with statistical and systematic uncertainties}

\end{table}
\endgroup

\begingroup
\squeezetable
\begin{table}[h]
\tabcolsep=0.11cm
\begin{tabular}{c|c|c}
&  $\sigma\ \  (cm^{2}/C^{12})\times 10^{-39}$ & $\sigma\ \  (cm^{2}/C^{12})\times 10^{-39}$ \\
$E_{\nu}$ \ (GeV)& neutrinos & antineutrinos  \\
\hline
$1.5 - 2.0$ & 3.952 $\pm$0.914 $\pm$0.454 &0.254 $\pm$0.494 $\pm$0.179 \\
$2.0 - 3.0$ & 2.485 $\pm$0.214 $\pm$0.335 &2.049 $\pm$0.290 $\pm$0.440\\
$3.0 - 4.0$ & 2.736 $\pm$0.183 $\pm$0.349 &2.673 $\pm$0.243 $\pm$0.417\\
$4.0 - 5.0$ & 3.096 $\pm$0.276 $\pm$0.559 &1.725 $\pm$0.311 $\pm$0.457\\
$5.0 - 7.0$ & 5.037 $\pm$0.478 $\pm$0.798 &3.928 $\pm$0.666 $\pm$0.763\\
$7.0 - 9.0$ & 6.540 $\pm$0.791 $\pm$0.844 &6.840 $\pm$1.294 $\pm$1.084\\
$9.0 - 11.0$ & 6.321 $\pm$1.033 $\pm$0.932 &9.917 $\pm$1.863 $\pm$1.601\\
$11.0 - 15.0$ & 8.205 $\pm$1.115 $\pm$1.149 &15.896 $\pm$2.245 $\pm$2.538\\
$15.0 - 20.0$ & 12.330 $\pm$1.686 $\pm$1.622 &9.931 $\pm$3.477 $\pm$1.768\\


\end{tabular}
\caption{The measurement of the cross section as a function of $E_\nu$ and
statistical and systematic uncertainties}
\end{table}
\endgroup

\begingroup
\squeezetable
\begin{table}[h]
\tabcolsep=0.11cm
\begin{tabular}{c|c|c|c|c|c|c|c|c|c}
$E_{\nu}(GeV)$ & $1.5 - 2.0$ & $2.0 - 3.0$ & $3.0 - 4.0$ & $4.0 - 5.0$ & $5.0 - 7.0$ & $7.0 - 9.0$ & $9.0 - 11.0$ & $11.0 - 15.0$ & $15.0 - 20.0$\\
\hline
$(\nu/cm^{2}/POT/ \times 10^{-8})$ &0.291	&0.865&	0.906&	0.375	&0.206&	0.100&	0.0664	&0.0871	&0.0404 \\
\hline
$({\bar{\nu}}/cm^{2}/POT/ \times 10^{-8})$ &0.265&	0.762&	0.754	&0.293&	0.140&	0.0556&	0.0329&	0.0391&	0.0154 \\
\end{tabular}
\caption{The predicted neutrino and anti-neutrino beam fluxes for the data included in this analysis}
\end{table}
\endgroup

\begingroup
\squeezetable
\begin{table}[h]
\tabcolsep=0.11cm
\begin{tabular}{c|ccccccccc}
$E_{\pi}(GeV)$ & 0 - 0.25 & 0.25 - 0.5 & 0.5 - 0.75 & 0.75 - 1.0 & 1.0 - 1.5 & 1.5 - 2.0 & 2.0 - 2.5 & 2.5 - 3.5 & 3.5 - 4 \\
\hline
0 - 0.25 &0.1356&	0.9767&	-0.1924&	-0.0664&	-0.0140&	-0.0009&	-0.0001&	-0.0000&	0.0000\\	
0.25 - 0.5 &0.9767&	0.4979&	0.0279&	-0.0191&	-0.0128&	-0.0031&	-0.0007&	-0.0001&	-0.0000	\\
0.5 - 0.75 &-0.1924&	0.0279&	0.1102&	0.0496&	0.0026&	-0.0041&	-0.0017&	-0.0004&	-0.0001	\\
0.75 - 1.0 &-0.0664&	-0.0191& 0.0496&	0.0920&	0.0353&	0.0005&	-0.0029&	-0.0011&	-0.0003	\\
1.0 - 1.5 &-0.0140&	-0.0128& 0.0026&	0.0353&	0.0603&	0.0174&	0.0016&	-0.0011&	-0.0008	\\
1.5 - 2.0 &-0.0009&	-0.0031& -0.0041&	0.0005&	0.0174&	0.0372&	0.0174&	0.0025&	-0.0007	\\
2.0 - 2.5 &-0.0001&	-0.0007&	-0.0017&	-0.0029&	0.0016&	0.0174&	0.0225&	0.0076&	0.0013	\\
2.5 - 3.5& -0.0000&	-0.0001&	-0.0004&	-0.0011&	-0.0011&	0.0025&	0.0076&	0.0088&	0.0037	\\
3.5 - 4 &0.0000	&-0.0000&	-0.0001&	-0.0003&	-0.0008&	-0.0007&	0.0013&	0.0037&	0.0037	\\
 
\end{tabular}
\caption{Neutrino $d\sigma/dE_\pi$ statistical covariance matrix  $\times 10^{-79}$.  Note that the full uncertainty of the result is obtained by adding this covariance matrix to those in Tables~\ref{tab:epi-flux} and \ref{tab:epi-nonflux}. }
\end{table}
\endgroup

\begingroup
\squeezetable
\begin{table}[h]
\tabcolsep=0.11cm
\begin{tabular}{c|ccccccccc}
$E_{\pi}(GeV)$ & 0 - 0.25 & 0.25 - 0.5 & 0.5 - 0.75 & 0.75 - 1.0 & 1.0 - 1.5 & 1.5 - 2.0 & 2.0 - 2.5 & 2.5 - 3.5 & 3.5 - 4 \\
\hline
0 - 0.25 &0.0469&	0.2898&	0.1137&	0.1111&	0.0805&	0.0392&	0.0194&	0.0092&	0.0046\\	
0.25 - 0.5& 0.2898&	1.7901&	0.7017&	0.6859&	0.4970&	0.2426&	0.1203&	0.0570&	0.0283	\\
0.5 - 0.75& 0.1137&	0.7017&	0.2757&	0.2696&	0.1950&	0.0946&	0.0466&	0.0221&	0.0111	\\
0.75 - 1.0& 0.1111&	0.6859&	0.2696&	0.2636&	0.1907&	0.0925&	0.0455&	0.0216&	0.0108	\\
1.0 - 1.5& 0.0805&	0.4970&	0.1950&	0.1907&	0.1381&	0.0673&	0.0333&	0.0158&	0.0079	\\
1.5 - 2.0& 0.0392&	0.2426&	0.0946&	0.0925&	0.0673&	0.0334&	0.0169&	0.0080&	0.0039	\\
2.0 - 2.5& 0.0194&	0.1203&	0.0466&	0.0455&	0.0333&	0.0169&	0.0086&	0.0041&	0.0020	\\
2.5 - 3.5& 0.0092&	0.0570&	0.0221&	0.0216&	0.0158&	0.0080&	0.0041&	0.0019&	0.0009	\\
3.5 - 4& 0.0046&	0.0283&	0.0111&	0.0108&	0.0079&	0.0039&	0.0020&	0.0009&	0.0005	\\
 
\end{tabular}
\caption{Neutrino $d\sigma/dE_\pi$ flux systematic covariance matrix  $\times 10^{-79}$ }
\label{tab:epi-flux}
\end{table}
\endgroup

\begingroup
\squeezetable
\begin{table}[h]
\tabcolsep=0.11cm
\begin{tabular}{c|ccccccccc}
$E_{\pi}(GeV)$ & 0 - 0.25 & 0.25 - 0.5 & 0.5 - 0.75 & 0.75 - 1.0 & 1.0 - 1.5 & 1.5 - 2.0 & 2.0 - 2.5 & 2.5 - 3.5 & 3.5 - 4 \\
\hline
0 - 0.25 &0.7835&	0.1193&	0.0369&	0.0302&	0.0170&	0.0085&	0.0056&	0.0029&	0.0012	\\
0.25 - 0.5& 0.1193&	1.4417&	0.2816&	0.1973&	0.0913&	0.0386&	0.0276&	0.0152&	0.0064	\\
0.5 - 0.75& 0.0369&	0.2816&	0.2735&	0.1186&	0.0631&	0.0243&	0.0129&	0.0062&	0.0030	\\
0.75 - 1.0 &0.0302&	0.1973&	0.1186&	0.2271&	0.0666&	0.0278&	0.0142&	0.0065&	0.0031	\\
1.0 - 1.5 &0.0170&	0.0913&	0.0631&	0.0666&	0.1210&	0.0205&	0.0107&	0.0047&	0.0021	\\
1.5 - 2.0 &0.0085&	0.0386&	0.0243&	0.0278&	0.0205&	0.0550&	0.0068&	0.0030&	0.0012	\\
2.0 - 2.5 &0.0056&	0.0276&	0.0129&	0.0142&	0.0107&	0.0068&	0.0277&	0.0021&	0.0008	\\
2.5 - 3.5 &0.0029&	0.0152&	0.0062&	0.0065&	0.0047&	0.0030&	0.0021&	0.0100&	0.0005\\	
3.5 - 4 &0.0012	&0.0064&	0.0030&	0.0031&	0.0021&	0.0012&	0.0008&	0.0005&	0.0037	\\
 
\end{tabular}
\caption{Neutrino $d\sigma/dE_\pi$ non-flux systematic covariance matrix  $\times 10^{-79}$ }
\label{tab:epi-nonflux}
\end{table}
\endgroup

\begingroup
\squeezetable
\begin{table}[h]
\tabcolsep=0.11cm
\begin{tabular}{c|ccccccccc}
$E_{\pi}(GeV)$ & 0 - 0.25 & 0.25 - 0.5 & 0.5 - 0.75 & 0.75 - 1.0 & 1.0 - 1.5 & 1.5 - 2.0 & 2.0 - 2.5 & 2.5 - 3.5 & 3.5 - 4 \\
\hline
0 - 0.25& 0.2141	&0.5504	&-0.1078	&-0.0621	&-0.0139	&-0.0004	&0.0001	&0.0001	&-0.0000	\\
0.25 - 0.5 &0.5504	&0.9801	&0.1167	&-0.0276	&-0.0291	&-0.0081	&-0.0018	&-0.0003&	-0.0001	\\
0.5 - 0.75 &-0.1078	&0.1167	&0.2017 &0.1025	&0.0074	&-0.0087	&-0.0034	&-0.0008	&-0.0001	\\
0.75 - 1.0 &-0.0621	&-0.0276	&0.1025	&0.1607	&0.0591	&0.0004	&-0.0052	&-0.0021	&-0.0005	\\
1.0 - 1.5 &-0.0139	&-0.0291	&0.0074	&0.0591	&0.0875	&0.0295	&0.0034	&-0.0016	&-0.0010	\\
1.5 - 2.0 &-0.0004	&-0.0081	&-0.0087	&0.0004	&0.0295	&0.0535	&0.0241	&0.0040	&-0.0005	\\
2.0 - 2.5 &0.0001	&-0.0018	&-0.0034	&-0.0052	&0.0034	&0.0241	&0.0345	&0.0125	&0.0018	\\
2.5 - 3.5 &0.0001	&-0.0003	&-0.0008	&-0.0021	&-0.0016	&0.0040	&0.0125	&0.0126	&0.0047	\\
3.5 - 4 &-0.0000	&-0.0001	&-0.0001	&-0.0005	&-0.0010	&-0.0005	&0.0018	&0.0047	&0.0028\\	
 
\end{tabular}
\caption{Anti-neutrino $d\sigma/dE_\pi$ statistical covariance matrix  $\times 10^{-79}$.  Note that the full uncertainty of the result is obtained by adding this covariance matrix to those in Tables~\ref{tab:epiB-flux} and \ref{tab:epiB-nonflux}. }
\end{table}
\endgroup

\begingroup
\squeezetable
\begin{table}[h]
\tabcolsep=0.11cm
\begin{tabular}{c|ccccccccc}
$E_{\pi}(GeV)$ & 0 - 0.25 & 0.25 - 0.5 & 0.5 - 0.75 & 0.75 - 1.0 & 1.0 - 1.5 & 1.5 - 2.0 & 2.0 - 2.5 & 2.5 - 3.5 & 3.5 - 4 \\
\hline
0 - 0.25 &0.0420	&0.1947	&0.1071	&0.1179	&0.0825	&0.0341	&0.0122	&0.0019	&0.0012	\\
0.25 - 0.5 &0.1947	&0.9108	&0.4947	&0.5428	&0.3779	&0.1540	&0.0537	&0.0074	&0.0049	\\
0.5 - 0.75 &0.1071	&0.4947	&0.2756	&0.3041	&0.2131	&0.0886	&0.0319	&0.0052	&0.0032	\\
0.75 - 1.0 &0.1179	&0.5428	&0.3041	&0.3361	&0.2360	&0.0987	&0.0358	&0.0060	&0.0036	\\
1.0 - 1.5 &0.0825	&0.3779	&0.2131	&0.2360	&0.1663	&0.0703	&0.0259	&0.0046	&0.0026	\\
1.5 - 2.0 &0.0341	&0.1540	&0.0886	&0.0987	&0.0703	&0.0306	&0.0118	&0.0025	&0.0013	\\
2.0 - 2.5 &0.0122	&0.0537	&0.0319	&0.0358	&0.0259	&0.0118	&0.0048	&0.0012	&0.0006	\\
2.5 - 3.5 &0.0019	&0.0074	&0.0052	&0.0060	&0.0046	&0.0025	&0.0012	&0.0005	&0.0002	\\
3.5 - 4 &0.0012	&0.0049	&0.0032	&0.0036	&0.0026	&0.0013	&0.0006	&0.0002	&0.0001\\	
 
\end{tabular}
\caption{Anti-neutrino $d\sigma/dE_\pi$ flux systematic covariance matrix  $\times 10^{-79}$ }
\label{tab:epiB-flux}
\end{table}
\endgroup

\begingroup
\squeezetable
\begin{table}[h]
\tabcolsep=0.11cm
\begin{tabular}{c|ccccccccc}
$E_{\pi}(GeV)$ & 0 - 0.25 & 0.25 - 0.5 & 0.5 - 0.75 & 0.75 - 1.0 & 1.0 - 1.5 & 1.5 - 2.0 & 2.0 - 2.5 & 2.5 - 3.5 & 3.5 - 4 \\
\hline
0 - 0.25& 0.5242	&0.6582	&0.1663	&0.0840	&0.0262	&0.0080	&0.0058	&0.0020	&0.0002	\\
0.25 - 0.5 &0.6582	&5.8193	&1.0820	&0.5440	&0.1697	&0.0509	&0.0383	&0.0153	&0.0027	\\
0.5 - 0.75 &0.1663	&1.0820	&0.7139	&0.2271	&0.0783	&0.0256	&0.0199	&0.0094	&0.0026	\\
0.75 - 1.0 &0.0840	&0.5440	&0.2271	&0.4017	&0.0557	&0.0196	&0.0145	&0.0070	&0.0021	\\
1.0 - 1.5 &0.0262	&0.1697	&0.0783	&0.0557	&0.1649	&0.0106	&0.0072	&0.0036	&0.0012	\\
1.5 - 2.0 &0.0080	&0.0509	&0.0256	&0.0196	&0.0106	&0.0903	&0.0051	&0.0025	&0.0008	\\
2.0 - 2.5 &0.0058	&0.0383	&0.0199	&0.0145	&0.0072	&0.0051	&0.0505	&0.0023	&0.0007	\\
2.5 - 3.5 &0.0020	&0.0153	&0.0094	&0.0070	&0.0036	&0.0025	&0.0023	&0.0195	&0.0005	\\
3.5 - 4 &0.0002	&0.0027	&0.0026	&0.0021	&0.0012	&0.0008	&0.0007	&0.0005	&0.0041\\	
 
\end{tabular}
\caption{Anti-neutrino $d\sigma/dE_\pi$ non-flux systematic covariance matrix  $\times 10^{-79}$ }
\label{tab:epiB-nonflux}
\end{table}
\endgroup

\begingroup
\squeezetable
\begin{table}[h]
\tabcolsep=0.11cm
\begin{tabular}{c|cccccccccccc}
Degrees & 0 - 5 & 5 - 10 & 10 -15 & 15 - 20 & 20 - 25 & 25 - 30 & 30 - 35 & 35 - 40 & 40 - 45 & 45 - 50 & 50 - 60 & 60 - 70 \\
\hline
 0 - 5& 0.1518 & 0.0052 & -0.0114 & -0.0053 & -0.0016 & -0.0015 & -0.0012 & -0.0008 & -0.0003 & -0.0001 & 0.0000 & 0.0000  \\
5 - 10 &0.0052 & 0.1256 & 0.0078 & -0.0083 & -0.0031 & -0.0012 & -0.0005 & -0.0003 & -0.0004 & -0.0001 & -0.0000 & -0.0000  \\
10 -15 &-0.0114 & 0.0078 & 0.0731 & 0.0085 & -0.0045 & -0.0018 & -0.0009 & -0.0004 & -0.0002 & -0.0001 & -0.0000 & -0.0000  \\
15 - 20  &-0.0053 & -0.0083 & 0.0085 & 0.0569 & 0.0067 & -0.0030 & -0.0015 & -0.0006 & 0.0002 & -0.0001 & -0.0001 & -0.0000  \\
20 - 25 &-0.0016 & -0.0031 & -0.0045 & 0.0067 & 0.0374 & 0.0056 & -0.0015 & -0.0005 & -0.0001 & 0.0002 & -0.0000 & -0.0000  \\
25 - 30 &-0.0015 & -0.0012 & -0.0018 & -0.0030 & 0.0056 & 0.0289 & 0.0050 & -0.0012 & -0.0028 & -0.0005 & 0.0000 & 0.0000  \\
30 - 35 &-0.0012 & -0.0005 & -0.0009 & -0.0015 & -0.0015 & 0.0050 & 0.0241 & 0.0054 & -0.0023 & -0.0024 & -0.0004 & 0.0000  \\
35 - 40 &-0.0008 & -0.0003 & -0.0004 & -0.0006 & -0.0005 & -0.0012 & 0.0054 & 0.0186 & 0.0072 & -0.0007 & -0.0011 & -0.0002  \\
40 - 45 &-0.0003 & -0.0004 & -0.0002 & 0.0002 & -0.0001 & -0.0028 & -0.0023 & 0.0072 & 0.0139 & 0.0058 & -0.0005 & -0.0005  \\
45 - 50 &-0.0001 & -0.0001 & -0.0001 & -0.0001 & 0.0002 & -0.0005 & -0.0024 & -0.0007 & 0.0058 & 0.0092 & 0.0024 & -0.0001  \\
50 - 60 &0.0000 & -0.0000 & -0.0000 & -0.0001 & -0.0000 & 0.0000 & -0.0004 & -0.0011 & -0.0005 & 0.0024 & 0.0057 & 0.0014  \\
60 - 70 &0.0000 & -0.0000 & -0.0000 & -0.0000 & -0.0000 & 0.0000 & 0.0000 & -0.0002 & -0.0005 & -0.0001 & 0.0014 & 0.0053  \\
 
\end{tabular}
\caption{Neutrino $d\sigma/d\theta_\pi$ statistical covariance matrix  $\times 10^{-81}$.  Note that the full uncertainty of the result is obtained by adding this covariance matrix to those in Tables~\ref{tab:thpi-flux} and \ref{tab:thpi-nonflux}. }
\end{table}
\endgroup

\begingroup
\squeezetable
\begin{table}[h]
\tabcolsep=0.11cm
\begin{tabular}{c|cccccccccccc}
Degrees & 0 - 5 & 5 - 10 & 10 -15 & 15 - 20 & 20 - 25 & 25 - 30 & 30 - 35 & 35 - 40 & 40 - 45 & 45 - 50 & 50 - 60 & 60 - 70 \\
\hline
 0 - 5&0.0844 & 0.0923 & 0.0849 & 0.0911 & 0.0671 & 0.0540 & 0.0479 & 0.0365 & 0.0290 & 0.0150 & 0.0086 & 0.0016 \\ 
5 - 10 &0.0923 & 0.1013 & 0.0931 & 0.0998 & 0.0735 & 0.0592 & 0.0524 & 0.0400 & 0.0318 & 0.0165 & 0.0095 & 0.0018  \\
10 -15 &0.0849 & 0.0931 & 0.0856 & 0.0918 & 0.0676 & 0.0544 & 0.0482 & 0.0368 & 0.0292 & 0.0152 & 0.0087 & 0.0016  \\
15 - 20 &0.0911 & 0.0998 & 0.0918 & 0.0985 & 0.0725 & 0.0584 & 0.0517 & 0.0394 & 0.0313 & 0.0163 & 0.0093 & 0.0017  \\
20 - 25 &0.0671 & 0.0735 & 0.0676 & 0.0725 & 0.0534 & 0.0430 & 0.0381 & 0.0290 & 0.0231 & 0.0120 & 0.0069 & 0.0013  \\
25 - 30 &0.0540 & 0.0592 & 0.0544 & 0.0584 & 0.0430 & 0.0346 & 0.0307 & 0.0234 & 0.0186 & 0.0096 & 0.0055 & 0.0010  \\
30 - 35 &0.0479 & 0.0524 & 0.0482 & 0.0517 & 0.0381 & 0.0307 & 0.0272 & 0.0207 & 0.0165 & 0.0085 & 0.0049 & 0.0009  \\
35 - 40 &0.0365 & 0.0400 & 0.0368 & 0.0394 & 0.0290 & 0.0234 & 0.0207 & 0.0158 & 0.0126 & 0.0065 & 0.0038 & 0.0007  \\
40 - 45 &0.0290 & 0.0318 & 0.0292 & 0.0313 & 0.0231 & 0.0186 & 0.0165 & 0.0126 & 0.0100 & 0.0052 & 0.0030 & 0.0006  \\
45 - 50 &0.0150 & 0.0165 & 0.0152 & 0.0163 & 0.0120 & 0.0096 & 0.0085 & 0.0065 & 0.0052 & 0.0027 & 0.0016 & 0.0003  \\
50 - 60 &0.0086 & 0.0095 & 0.0087 & 0.0093 & 0.0069 & 0.0055 & 0.0049 & 0.0038 & 0.0030 & 0.0016 & 0.0009 & 0.0002  \\
60 - 70 &0.0016 & 0.0018 & 0.0016 & 0.0017 & 0.0013 & 0.0010 & 0.0009 & 0.0007 & 0.0006 & 0.0003 & 0.0002 & 0.0001  \\
 
\end{tabular}
\caption{Neutrino $d\sigma/d\theta_\pi$ flux systematic covariance matrix  $\times 10^{-81}$ }
\label{tab:thpi-flux}
\end{table}
\endgroup

\begingroup
\squeezetable
\begin{table}[h]
\tabcolsep=0.11cm
\begin{tabular}{c|cccccccccccc}
Degrees & 0 - 5 & 5 - 10 & 10 -15 & 15 - 20 & 20 - 25 & 25 - 30 & 30 - 35 & 35 - 40 & 40 - 45 & 45 - 50 & 50 - 60 & 60 - 70 \\
\hline
0 - 5&0.1420 & 0.1742 & 0.1132 & 0.0416 & 0.0509 & 0.0353 & 0.0573 & 0.0565 & 0.0422 & 0.0258 & 0.0460 & 0.0530  \\
5 - 10 &0.1742 & 0.2246 & 0.1452 & 0.0569 & 0.0686 & 0.0492 & 0.0748 & 0.0740 & 0.0559 & 0.0351 & 0.0603 & 0.0682  \\
10 -15 &0.1132 & 0.1452 & 0.1047 & 0.0389 & 0.0476 & 0.0350 & 0.0532 & 0.0524 & 0.0399 & 0.0254 & 0.0420 & 0.0466  \\
15 - 20 &0.0416 & 0.0569 & 0.0389 & 0.0289 & 0.0235 & 0.0192 & 0.0218 & 0.0208 & 0.0157 & 0.0099 & 0.0147 & 0.0155  \\
20 - 25 &0.0509 & 0.0686 & 0.0476 & 0.0235 & 0.0380 & 0.0223 & 0.0277 & 0.0267 & 0.0204 & 0.0131 & 0.0194 & 0.0203  \\
25 - 30 &0.0353 & 0.0492 & 0.0350 & 0.0192 & 0.0223 & 0.0299 & 0.0218 & 0.0210 & 0.0163 & 0.0109 & 0.0149 & 0.0146  \\
30 - 35 &0.0573 & 0.0748 & 0.0532 & 0.0218 & 0.0277 & 0.0218 & 0.0433 & 0.0332 & 0.0261 & 0.0179 & 0.0262 & 0.0267  \\
35 - 40 &0.0565 & 0.0740 & 0.0524 & 0.0208 & 0.0267 & 0.0210 & 0.0332 & 0.0425 & 0.0265 & 0.0185 & 0.0271 & 0.0273  \\
40 - 45 &0.0422 & 0.0559 & 0.0399 & 0.0157 & 0.0204 & 0.0163 & 0.0261 & 0.0265 & 0.0279 & 0.0155 & 0.0224 & 0.0220  \\
45 - 50 &0.0258 & 0.0351 & 0.0254 & 0.0099 & 0.0131 & 0.0109 & 0.0179 & 0.0185 & 0.0155 & 0.0171 & 0.0167 & 0.0157  \\
50 - 60 &0.0460 & 0.0603 & 0.0420 & 0.0147 & 0.0194 & 0.0149 & 0.0262 & 0.0271 & 0.0224 & 0.0167 & 0.0298 & 0.0250  \\
60 - 70 &0.0530 & 0.0682 & 0.0466 & 0.0155 & 0.0203 & 0.0146 & 0.0267 & 0.0273 & 0.0220 & 0.0157 & 0.0250 & 0.0275  \\
\end{tabular}
\caption{Neutrino $d\sigma/d\theta_\pi$ non-flux systematic covariance matrix  $\times 10^{-81}$ }
\label{tab:thpi-nonflux}
\end{table}
\endgroup

\begingroup
\squeezetable
\begin{table}[h]
\tabcolsep=0.11cm
\begin{tabular}{c|cccccccccccc}
Degrees & 0 - 5 & 5 - 10 & 10 -15 & 15 - 20 & 20 - 25 & 25 - 30 & 30 - 35 & 35 - 40 & 40 - 45 & 45 - 50 & 50 - 60 & 60 - 70 \\
\hline
0 - 5 &0.1863 & 0.0072 & -0.0147 & -0.0066 & -0.0021 & -0.0017 & -0.0015 & -0.0007 & -0.0004 & -0.0004 & -0.0002 & -0.0000  \\
5 - 10 &0.0072 & 0.2111 & 0.0400 & 0.0327 & 0.0796 & 0.0851 & 0.1135 & 0.0819 & -0.0005 & -0.0002 & -0.0001 & -0.0000  \\
10 -15 &-0.0147 & 0.0400 & 0.1386 & 0.0409 & 0.0420 & 0.0489 & 0.0665 & 0.0484 & -0.0007 & -0.0003 & -0.0000 & 0.0000  \\
15 - 20 &-0.0066 & 0.0327 & 0.0409 & 0.1249 & 0.0720 & 0.0580 & 0.0802 & 0.0588 & -0.0002 & -0.0004 & -0.0001 & -0.0000  \\
20 - 25 &-0.0021 & 0.0796 & 0.0420 & 0.0720 & 0.1741 & 0.1116 & 0.1304 & 0.0950 & 0.0001 & 0.0003 & -0.0001 & -0.0001  \\
25 - 30 &-0.0017 & 0.0851 & 0.0489 & 0.0580 & 0.1116 & 0.1365 & 0.1258 & 0.0823 & -0.0050 & -0.0008 & 0.0000 & 0.0000 \\
30 - 35 &-0.0015 & 0.1135 & 0.0665 & 0.0802 & 0.1304 & 0.1258 & 0.1785 & 0.1090 & -0.0053 & -0.0047 & -0.0006 & 0.0001  \\
35 - 40 &-0.0007 & 0.0819 & 0.0484 & 0.0588 & 0.0950 & 0.0823 & 0.1090 & 0.1135 & 0.0152 & -0.0014 & -0.0024 & -0.0004  \\
40 - 45 &-0.0004 & -0.0005 & -0.0007 & -0.0002 & 0.0001 & -0.0050 & -0.0053 & 0.0152 & 0.0314 & 0.0137 & -0.0018 & -0.0013\\  
45 - 50 &-0.0004 & -0.0002 & -0.0003 & -0.0004 & 0.0003 & -0.0008 & -0.0047 & -0.0014 & 0.0137 & 0.0198 & 0.0042 & -0.0005  \\
50 - 60 &-0.0002 & -0.0001 & -0.0000 & -0.0001 & -0.0001 & 0.0000 & -0.0006 & -0.0024 & -0.0018 & 0.0042 & 0.0123 & 0.0028  \\
60 - 7 &-0.0000 & -0.0000 & 0.0000 & -0.0000 & -0.0001 & 0.0000 & 0.0001 & -0.0004 & -0.0013 & -0.0005 & 0.0028 & 0.0119 \\
  
\end{tabular}
\caption{Anti-neutrino $d\sigma/d\theta_\pi$ statistical covariance matrix  $\times 10^{-81}$.  Note that the full uncertainty of the result is obtained by adding this covariance matrix to those in Tables~\ref{tab:thpiB-flux} and \ref{tab:thpiB-nonflux}. }
\end{table}
\endgroup

\begingroup
\squeezetable
\begin{table}[h]
\tabcolsep=0.11cm
\begin{tabular}{c|cccccccccccc}
Degrees & 0 - 5 & 5 - 10 & 10 -15 & 15 - 20 & 20 - 25 & 25 - 30 & 30 - 35 & 35 - 40 & 40 - 45 & 45 - 50 & 50 - 60 & 60 - 70 \\
\hline
 0 - 5&0.0552 & 0.0801 & 0.0547 & 0.0661 & 0.0640 & 0.0369 & 0.0322 & 0.0227 & 0.0184 & 0.0014 & -0.0006 & 0.0042  \\
 5 - 10 &0.0801 & 0.1166 & 0.0797 & 0.0961 & 0.0932 & 0.0540 & 0.0470 & 0.0333 & 0.0269 & 0.0022 & -0.0008 & 0.0061  \\
10 -15 &0.0547 & 0.0797 & 0.0547 & 0.0659 & 0.0641 & 0.0370 & 0.0323 & 0.0229 & 0.0187 & 0.0016 & -0.0004 & 0.0043  \\
15 - 20 &0.0661 & 0.0961 & 0.0659 & 0.0796 & 0.0774 & 0.0446 & 0.0390 & 0.0277 & 0.0226 & 0.0019 & -0.0005 & 0.0052  \\
20 - 25 &0.0640 & 0.0932 & 0.0641 & 0.0774 & 0.0760 & 0.0438 & 0.0385 & 0.0276 & 0.0228 & 0.0022 & -0.0003 & 0.0052  \\
25 - 30 &0.0369 & 0.0540 & 0.0370 & 0.0446 & 0.0438 & 0.0255 & 0.0223 & 0.0160 & 0.0132 & 0.0013 & -0.0001 & 0.0030  \\
30 - 35 &0.0322 & 0.0470 & 0.0323 & 0.0390 & 0.0385 & 0.0223 & 0.0196 & 0.0141 & 0.0117 & 0.0012 & -0.0001 & 0.0027  \\
35 - 40 &0.0227 & 0.0333 & 0.0229 & 0.0277 & 0.0276 & 0.0160 & 0.0141 & 0.0102 & 0.0086 & 0.0010 & 0.0000 & 0.0019 \\
40 - 45 &0.0184 & 0.0269 & 0.0187 & 0.0226 & 0.0228 & 0.0132 & 0.0117 & 0.0086 & 0.0074 & 0.0009 & 0.0002 & 0.0017  \\
45 - 50 &0.0014 & 0.0022 & 0.0016 & 0.0019 & 0.0022 & 0.0013 & 0.0012 & 0.0010 & 0.0009 & 0.0002 & 0.0001 & 0.0002  \\
50 - 60 &-0.0006 & -0.0008 & -0.0004 & -0.0005 & -0.0003 & -0.0001 & -0.0001 & 0.0000 & 0.0002 & 0.0001 & 0.0001 & 0.0001 \\  
60 - 70 &0.0042 & 0.0061 & 0.0043 & 0.0052 & 0.0052 & 0.0030 & 0.0027 & 0.0019 & 0.0017 & 0.0002 & 0.0001 & 0.0004  \\
 
\end{tabular}
\caption{Anti-neutrino $d\sigma/d\theta_\pi$ flux systematic covariance matrix  $\times 10^{-81}$ }
\label{tab:thpiB-flux}
\end{table}
\endgroup

\clearpage  

\begingroup
\squeezetable
\begin{table}[h]
\tabcolsep=0.11cm
\begin{tabular}{c|cccccccccccc}
Degrees & 0 - 5 & 5 - 10 & 10 -15 & 15 - 20 & 20 - 25 & 25 - 30 & 30 - 35 & 35 - 40 & 40 - 45 & 45 - 50 & 50 - 60 & 60 - 70 \\
\hline
 0 - 5&0.0931 & 0.0923 & 0.0859 & 0.0256 & 0.0682 & 0.0426 & 0.0683 & 0.0556 & 0.0543 & 0.0173 & 0.0259 & 0.0859 \\ 
 5 - 10&0.0923 & 0.1068 & 0.0968 & 0.0311 & 0.0804 & 0.0525 & 0.0815 & 0.0675 & 0.0671 & 0.0246 & 0.0319 & 0.0988 \\
10 -15 &0.0859 & 0.0968 & 0.1070 & 0.0336 & 0.0804 & 0.0567 & 0.0837 & 0.0692 & 0.0696 & 0.0300 & 0.0353 & 0.0917  \\
15 - 20 &0.0256 & 0.0311 & 0.0336 & 0.0280 & 0.0347 & 0.0295 & 0.0354 & 0.0330 & 0.0334 & 0.0220 & 0.0182 & 0.0305  \\
20 - 25 &0.0682 & 0.0804 & 0.0804 & 0.0347 & 0.0848 & 0.0550 & 0.0756 & 0.0663 & 0.0659 & 0.0331 & 0.0341 & 0.0807  \\
25 - 30 &0.0426 & 0.0525 & 0.0567 & 0.0295 & 0.0550 & 0.0545 & 0.0566 & 0.0514 & 0.0519 & 0.0313 & 0.0280 & 0.0524  \\
30 - 35 &0.0683 & 0.0815 & 0.0837 & 0.0354 & 0.0756 & 0.0566 & 0.0856 & 0.0678 & 0.0682 & 0.0349 & 0.0352 & 0.0803  \\
35 - 40 &0.0556 & 0.0675 & 0.0692 & 0.0330 & 0.0663 & 0.0514 & 0.0678 & 0.0690 & 0.0612 & 0.0339 & 0.0321 & 0.0674  \\
40 - 45 &0.0543 & 0.0671 & 0.0696 & 0.0334 & 0.0659 & 0.0519 & 0.0682 & 0.0612 & 0.0693 & 0.0349 & 0.0326 & 0.0664  \\
45 - 50 &0.0173 & 0.0246 & 0.0300 & 0.0220 & 0.0331 & 0.0313 & 0.0349 & 0.0339 & 0.0349 & 0.0310 & 0.0197 & 0.0248  \\
50 - 60 &0.0259 & 0.0319 & 0.0353 & 0.0182 & 0.0341 & 0.0280 & 0.0352 & 0.0321 & 0.0326 & 0.0197 & 0.0216 & 0.0321  \\
60 - 70 &0.0859 & 0.0988 & 0.0917 & 0.0305 & 0.0807 & 0.0524 & 0.0803 & 0.0674 & 0.0664 & 0.0248 & 0.0321 & 0.1003  \\
\end{tabular}
\caption{Anti-neutrino $d\sigma/d\theta_\pi$ non-flux systematic covariance matrix  $\times 10^{-81}$ }
\label{tab:thpiB-nonflux}
\end{table}
\endgroup

\begingroup
\squeezetable
\begin{table}[h]
\tabcolsep=0.11cm
\begin{tabular}{c|ccccccccc}
$E_{\nu}(GeV)$ & 1.5 - 2.0 & 2.0 - 3.0 & 3.0 - 4.0 & 4.0 - 5.0 & 5.0 - 7.0 & 7.0 - 9.0 & 9.0 - 11.0 & 11.0 - 15.0 & 15.0 - 20.0\\
\hline
1.5 - 2.0&0.8354	&-0.0235	&-0.1095	&-0.0082	&-0.0004	&0.0000	&0.0000	&-0.0000	&-0.0000\\	
2.0 - 3.0&-0.0235	&0.0460	&-0.0003	&-0.0029	&-0.0004	&-0.0000&	-0.0000	&-0.0000	&-0.0000	\\
3.0 - 4.0&-0.1095	&-0.0003	&0.0334	&0.0018	&-0.0016	&-0.0002	&-0.0000	&-0.0000	&-0.0000	\\
4.0 - 5.0&-0.0082	&-0.0029	&0.0018	&0.0762	&0.0054&	-0.0029	&-0.0004	&-0.0001	&-0.0000	\\
 5.0 - 7.0&-0.0004	&-0.0004	&-0.0016	&0.0054	&0.2284&	0.0100	&-0.0079	&-0.0031	&-0.0003	\\
7.0 - 9.0&0.0000	&-0.0000	&-0.0002	&-0.0029	&0.0100&	0.6263	&0.0793	&-0.0416	&-0.0109	\\
9.0 - 11.0&0.0000	&-0.0000&	-0.0000	&-0.0004	&-0.0079	&0.0793	&1.0676	&0.2118	&-0.0724	\\
11.0 - 15.0 &-0.0000	&-0.0000	&-0.0000	&-0.0001	&-0.0031	&-0.0416	&0.2118	&1.2436	&0.0420	\\
15.0 - 20.0&-0.0000	&-0.0000	&-0.0000	&-0.0000	&-0.0003	&-0.0109	&-0.0724	&0.0420&	2.8417	\\

\end{tabular}
\caption{Neutrino $\sigma(E_\nu)$ statistical covariance matrix  $\times 10^{-78}$.  Note that the full uncertainty of the result is obtained by adding this covariance matrix to those in Tables~\ref{tab:enu-flux} and \ref{tab:enu-nonflux}. }
\end{table}
\endgroup

\begingroup
\squeezetable
\begin{table}[h]
\tabcolsep=0.11cm
\begin{tabular}{c|ccccccccc}
$E_{\nu}(GeV)$ & 1.5 - 2.0 & 2.0 - 3.0 & 3.0 - 4.0 & 4.0 - 5.0 & 5.0 - 7.0 & 7.0 - 9.0 & 9.0 - 11.0 & 11.0 - 15.0 & 15.0 - 20.0\\
\hline
1.5 - 2.0 &0.1607	&0.1059	&0.0963	&0.1011	&0.2184	&0.3017	&0.3280	&0.4243	&0.6208\\	
2.0 - 3.0  &0.1059	&0.0721	&0.0614	&0.0512	&0.1267	&0.1993	&0.2206	&0.2857	&0.4143	\\
3.0 - 4.0 &0.0963	&0.0614	&0.0727	&0.1065	&0.1662	&0.1811	&0.1884	&0.2465	&0.3571	\\
4.0 - 5.0 &0.1011	&0.0512	&0.1065	&0.2545	&0.3247	&0.1929	&0.1667	&0.2216	&0.3363	\\
5.0 - 7.0 &0.2184	&0.1267	&0.1662	&0.3247	&0.5415	&0.4320	&0.4100	&0.5299	&0.7969	\\
 7.0 - 9.0 &0.3017	&0.1993	&0.1811	&0.1929	&0.4320	&0.5928	&0.6354	&0.8159	&1.1904	\\
9.0 - 11.0 &0.3280	&0.2206	&0.1884	&0.1667	&0.4100	&0.6354	&0.6998	&0.9011	&1.3109	\\
11.0 - 15.0 &0.4243	&0.2857	&0.2465	&0.2216	&0.5299	&0.8159	&0.9011	&1.1661	&1.6954	\\
15.0 - 20.0 &0.6208	&0.4143	&0.3571	&0.3363	&0.7969	&1.1904	&1.3109	&1.6954	&2.4803	\\
\end{tabular}
\caption{Neutrino $\sigma(E_\nu)$  flux systematic covariance matrix  $\times 10^{-78}$ }
\label{tab:enu-flux}
\end{table}
\endgroup

\begingroup
\squeezetable
\begin{table}[h]
\tabcolsep=0.11cm
\begin{tabular}{c|ccccccccc}
$E_{\nu}(GeV)$ & 1.5 - 2.0 & 2.0 - 3.0 & 3.0 - 4.0 & 4.0 - 5.0 & 5.0 - 7.0 & 7.0 - 9.0 & 9.0 - 11.0 & 11.0 - 15.0 & 15.0 - 20.0\\
\hline
1.5 - 2.0& 0.0454	&0.0315	&0.0316	&0.0320	&0.0366	&0.0440	&0.0506	&0.0439	&0.0410	\\
2.0 - 3.0 &0.0315	&0.0401	&0.0312	&0.0320	&0.0374	&0.0444	&0.0510	&0.0469	&0.0425	\\
3.0 - 4.0 & 0.0316	&0.0312	&0.0489	&0.0339	&0.0404	&0.0486	&0.0566	&0.0514	&0.0448	\\
4.0 - 5.0 &0.0320	&0.0320	&0.0339	&0.0577	&0.0420	&0.0503	&0.0592	&0.0545	&0.0479	\\
5.0 - 7.0 &0.0366	&0.0374	&0.0404	&0.0420	&0.0961	&0.0620	&0.0735	&0.0710	&0.0627	\\
7.0 - 9.0 &0.0440	&0.0444	&0.0486	&0.0503	&0.0620	&0.1203	&0.0906	&0.0830	&0.0720	\\
9.0 - 11.0 &0.0506	&0.0510	&0.0566	&0.0592	&0.0735	&0.0906	&0.1689	&0.1005	&0.0847	\\
11.0 - 15.0 &0.0439	&0.0469	&0.0514	&0.0545	&0.0710	&0.0830	&0.1005	&0.1551	&0.0942	\\
15.0 - 20.0 &0.0410	&0.0425	&0.0448	&0.0479	&0.0627	&0.0720	&0.0847	&0.0942	&0.1494	\\
  
\end{tabular}
\caption{Neutrino $\sigma(E_\nu)$ non-flux systematic covariance matrix  $\times 10^{-78}$ }
\label{tab:enu-nonflux}
\end{table}
\endgroup

\begingroup
\squeezetable
\begin{table}[h]
\tabcolsep=0.11cm
\begin{tabular}{c|ccccccccc}
$E_{\nu}(GeV)$ & 1.5 - 2.0 & 2.0 - 3.0 & 3.0 - 4.0 & 4.0 - 5.0 & 5.0 - 7.0 & 7.0 - 9.0 & 9.0 - 11.0 & 11.0 - 15.0 & 15.0 - 20.0\\
\hline
 1.5 - 2.0 &  0.2440	&0.0194	&-0.0495	&-0.0014	&0.0044	&-0.0004	&-0.0000	&-0.0000&	0.0000	\\
2.0 - 3.0 &0.0194	&0.0838	&0.0015	&-0.0059	&-0.0009	&-0.0000	&0.0000	&0.0000	&-0.0000	\\
3.0 - 4.0 &-0.0495	&0.0015	&0.0590	&0.0059	&-0.0026	&-0.0005	&-0.0000	&-0.0000	&0.0000	\\
 4.0 - 5.0 &-0.0014	&-0.0059	&0.0059	&0.0967	&0.0159	&-0.0041	&-0.0010	&-0.0001	&-0.0000\\	
 5.0 - 7.0 &0.0044	&-0.0009	&-0.0026	&0.0159	&0.4430	&0.0274	&-0.0202	&-0.0057	&-0.0000\\	
7.0 - 9.0 &-0.0004	&-0.0000	&-0.0005	&-0.0041	&0.0274	&1.6737	&0.3357	&-0.1557	&-0.0181\\	
9.0 - 11.0 &-0.0000	&0.0000	&-0.0000	&-0.0010	&-0.0202	&0.3357	&3.4695	&0.5463	&-0.1972	\\
11.0 - 15.0 &-0.0000	&0.0000	&-0.0000	&-0.0001	&-0.0057	&-0.1557	&0.5463	&5.0386	&0.1996	\\
15.0 - 20.0& 0.0000	&-0.0000&	0.0000	&-0.0000	&-0.0000	&-0.0181	&-0.1972	&0.1996	&12.0922	\\
\end{tabular}
\caption{Anti-neutrino $\sigma(E_\nu)$ statistical covariance matrix  $\times 10^{-78}$.  Note that the full uncertainty of the result is obtained by adding this covariance matrix to those in Tables~\ref{tab:enuB-flux} and \ref{tab:enuB-nonflux}. }
\end{table}
\endgroup

\begingroup
\squeezetable
\begin{table}[h]
\tabcolsep=0.11cm
\begin{tabular}{c|ccccccccc}
$E_{\nu}(GeV)$ & 1.5 - 2.0 & 2.0 - 3.0 & 3.0 - 4.0 & 4.0 - 5.0 & 5.0 - 7.0 & 7.0 - 9.0 & 9.0 - 11.0 & 11.0 - 15.0 & 15.0 - 20.0\\
\hline
1.5 - 2.0 & 0.0030	&0.0131	&0.0065	&-0.0044&0.0083	&0.0343	&0.0521	&0.0787	&0.0342	\\
 2.0 - 3.0 &0.0131	&0.0761	&0.0580	&0.0073	&0.0859	&0.2405	&0.3736	&0.6081	&0.3101	\\
3.0 - 4.0 &0.0065	&0.0580	&0.0721	&0.0565	&0.1320	&0.2336	&0.3659	&0.6337	&0.3403	\\
4.0 - 5.0 &-0.0044	&0.0073	&0.0565	&0.1085	&0.1740	&0.1277	&0.1769	&0.3409	&0.1411	\\
5.0 - 7.0 &0.0083	&0.0859	&0.1320	&0.1740	&0.4139	&0.4506	&0.6153	&1.0507	&0.4100	\\
7.0 - 9.0 &0.0343	&0.2405	&0.2336	&0.1277	&0.4506	&0.9108	&1.3782	&2.2888	&1.1480	\\
9.0 - 11.0 &0.0521	&0.3736	&0.3659	&0.1769	&0.6153	&1.3782	&2.1542	&3.6281	&1.9565	\\
11.0 - 15.0 &0.0787	&0.6081	&0.6337	&0.3409	&1.0507	&2.2888	&3.6281	&6.2100	&3.4856	\\
15.0 - 20.0& 0.0342	&0.3101	&0.3403	&0.1411	&0.4100	&1.1480	&1.9565	&3.4856	&2.3443	\\
\end{tabular}
\caption{Anti-neutrino $\sigma(E_\nu)$ flux systematic covariance matrix  $\times 10^{-78}$ }
\label{tab:enuB-flux}
\end{table}
\endgroup

\begingroup
\squeezetable
\begin{table}[h]
\tabcolsep=0.11cm
\begin{tabular}{c|ccccccccc}
$E_{\nu}(GeV)$ & 1.5 - 2.0 & 2.0 - 3.0 & 3.0 - 4.0 & 4.0 - 5.0 & 5.0 - 7.0 & 7.0 - 9.0 & 9.0 - 11.0 & 11.0 - 15.0 & 15.0 - 20.0\\
\hline
1.5 - 2.0 &0.0290	&0.0236	&0.0235	&0.0204	&0.0269	&0.0299	&0.0323	&0.0330	&0.0518	\\
2.0 - 3.0 &0.0236	&0.1172	&0.0995	&0.0951	&0.1161	&0.1354	&0.1505	&0.1069	&0.2238	\\
 3.0 - 4.0 &0.0235	&0.0995	&0.1021	&0.0900	&0.1115	&0.1298	&0.1450	&0.1098	&0.2149	\\
4.0 - 5.0 &0.0204	&0.0951	&0.0900	&0.1001	&0.1077	&0.1253	&0.1438	&0.1059	&0.2070	\\
5.0 - 7.0 &0.0269	&0.1161	&0.1115	&0.1077	&0.1689	&0.1719	&0.1970	&0.1474	&0.2863	\\
7.0 - 9.0 &0.0299	&0.1354	&0.1298	&0.1253	&0.1719	&0.2637	&0.2278	&0.1709	&0.3551	\\
9.0 - 11.0 &0.0323	&0.1505	&0.1450	&0.1438	&0.1970	&0.2278	&0.4093	&0.1736	&0.3761	\\
11.0 - 15.0 &0.0330	&0.1069	&0.1098	&0.1059	&0.1474	&0.1709	&0.1736	&0.2298	&0.2809	\\
15.0 - 20.0&0.0518	&0.2238	&0.2149	&0.2070	&0.2863	&0.3551	&0.3761	&0.2809	&0.7830	\\
  
\end{tabular}
\caption{Anti-neutrino $\sigma(E_\nu)$ non-flux systematic covariance matrix  $\times 10^{-78}$ }
\label{tab:enuB-nonflux}
\end{table}
\endgroup

\fi

\end{document}